\documentclass[11pt]{article}
\usepackage{graphicx}

\newcommand{\BABARPubYear}    {04}

\newcommand{\BABARConfNumber} {31}
\newcommand{\SLACPubNumber} {10612}

\usepackage{color}
\usepackage{graphics, graphicx}
\DeclareGraphicsExtensions{.eps}
\usepackage{epsfig}	

\definecolor{purple}{rgb}{0.5,0,0.5}
\definecolor{darkgreen}{rgb}{0.0,0.7,0.2}
\definecolor{forest}{rgb}{0.1,0.3,0.2}
\definecolor{darkred}{rgb}{0.7,0.0,0.2}
\definecolor{darkcyan}{rgb}{0.,0.2,0.8}
\definecolor{orange}{rgb}{0.7,0.2,0.0}

\newcommand{\recow}{\mbox{${w}$}}
\newcommand{\ctl}{\mbox{${\cos \theta_\ell}$}}
\newcommand{\ctv}{\mbox{${\cos \theta_V}$}}
\newcommand{\angchi}{\mbox{${\chi}$}}

\newcommand{\ffA}{\mbox{$A$}}

\newcommand{\ffH}{\mbox{$H$}}

\newcommand{\stl}{\mbox{${\sin \theta_l}$}}
\newcommand{\stv}{\mbox{${\sin \theta_V}$}}

\newcommand{\rone}{ {\mbox{$R_1$}}}
\newcommand{\rtwo}{{\mbox{$R_2$}}}

\def\rhosq{{\mbox{$\rho^2$}}}

\def\babar{BaBar}

\def\comb{combinatoric}

\def\sbr{sideband region}

\def\coef{coefficient}

\def\bkgd{background}

\def\dist{distribution}

\def\statl{statistical}

\def\syst{systematic}
\def\Syst{Systematic}

\def\param{parameter}

\def\recod{reconstructed}

\def\normzn{normalization}
\def\FF{form factor}

\def\FFcap{Form Factor}

\def\recon{reconstruction}

\def\bef{\begin{figure}}
\def\eef{\end{figure}}

\newcommand{\bsl}{\begin{slide}}
\newcommand{\esl}{\end{slide}}

\def\dstarlnu{\mbox{$D^{*}\ell \nu$}}

\def\Bbartodstarlnu{\mbox{${\bar B}^{0}\rightarrow D^{*+}\ell^{-} \nubar$}}

\def\bbartodstarlnu{\mbox{${ \bar{B}^0 \rightarrow D^* \ell  \nu}$ } }
\def\btodstarlnu{\mbox{${ B^0 \rightarrow D^* \ell  \nu}$ } }

\def\h0{H_{0}(q^2)}

\def\mdstar{M_{D^{*}}}

\def\cosby{\mbox{$\cos\theta_{BY}$}}

\def\m2miss{M^{2}_{\rm miss}}

\def\mb{M_B}

\def\momdstr{\wp_{D^{*}}}

\def\vdstr{v_{D^{*}}}

\def\upsfs{\mbox{$\Upsilon(4{\rm S})$}}

\def\be{\begin{equation}}
\def\ee{\end{equation}}
\def\bea{\begin{eqnarray}}
\def\eea{\end{eqnarray}}

\newcommand{\rp}{\right)}
\newcommand{\lp}{\left(}
\newcommand{\rpar}{\right)}
\newcommand{\lpar}{\left(}
\newcommand{\lb}{\left[}
\newcommand{\rb}{\right]}

\newcommand{\lbrackt}{\left[}
\newcommand{\rbrackt}{\right]}
\newcommand{\rbracket}{\right]}
\newcommand{\lbracket}{\left[}

\newbox\charbox
\newbox\slabox
\def\s#1{{      
        \setbox\charbox=\hbox{$#1$}
        \setbox\slabox=\hbox{$/$}
        \dimen\charbox=\ht\slabox
        \advance\dimen\charbox by -\dp\slabox
        \advance\dimen\charbox by -\ht\charbox
        \advance\dimen\charbox by \dp\charbox
        \divide\dimen\charbox by 2
        \raise-\dimen\charbox\hbox to \wd\charbox{\hss/\hss}
        \llap{$#1$}
}}

\newcommand{\as}{\mbox{$\alpha_{s}\ $}}
\newcommand{\asq}{\mbox{$\alpha_{s}^2\ $}}

 \newcommand{\benn}{\begin{displaymath}}
 \newcommand{\eenn}{\end{displaymath}}

 \newcommand{\bes}{\begin{displaymath}}
 \newcommand{\ees}{\end{displaymath}}

 \newcommand{\beas}{\begin{eqnarray*}}
 \newcommand{\eeas}{\end{eqnarray*}}

\newcommand{\bean}{\begin{eqnarray*}}
 \newcommand{\eean}{\end{eqnarray*}}

\newcommand{\rar}{\mbox{$\rightarrow$}}

\newcommand{\eminus}{\mbox{$e^-$}}
\newcommand{\eplus}{\mbox{$e^+$}}

\newcommand{\vub}{\mbox{$V_{ub}$}}

\newcommand{\vcb}{\mbox{$V_{cb}$}}

\newcommand{\nubar}{\overline{\nu}}

\newcommand{\dstar}{\mbox{$D^{*}$}}

\newcommand{\Bbar}{\overline{B}}

\newcommand{\Dbar}{\overline{D}}

\newcommand{\bi}{\begin{itemize}}
\newcommand{\ei}{\end{itemize}}

\newcommand{\Btodslnu}{\mbox{ $B \rar D^* \ell \nu$} }

\newcommand{\rhosqaone}{\rho^2_{A_1}}

\newcommand{\ifb}{\mbox{ $\rm fb^{-1}$}}

\newcommand{\Dzero}{\mbox{$D^0$}}

\newcommand{\Dstar}{\mbox{$D^*$}}
\newcommand{\Dstarplus}{\mbox{$D^{*+}$}}

\newcommand{\Dst}{\mbox{ $D^*$} }

\newcommand{\deltam}{\mbox{$\delta m$}}

\newcommand{\vsp}{\vspace{.5cm}}
\newcommand{\vsps}{\vspace{.5cm}}

\def\kv{kinematic variable}

\newcommand{\Bz}{$B^0$}

\def\lqcdomx{ { \lp {\Lambda_{\rm QCD} \over m_x }\rp  }}

\newcommand{\defineCat}[1]{%
\expandafter\def\csname #1\endcsname{{\ttfamily #1}\xspace}}
\defineCat{svtDch}
\defineCat{Dmode}
\defineCat{tagCat}
\defineCat{angCut}
\defineCat{leptID}
\defineCat{onOffRes}
\defineCat{peakGroup}
\defineCat{yields}

\def\resln{resolution}

\def\ssec{\subsection}

\def\bcenter{\begin{center}}
\def\ecenter{\end{center}}

\def\approxx{$\sim$}

\def\Appx{Appendix}

\def\haone{h_{A_1}}

\def\dstarell{\mbox{$D^*-\ell$}}

\RequirePackage{xspace}

\usepackage{relsize}
\def\babar{\mbox{\slshape B\kern-0.1em{\smaller A}\kern-0.1em
    B\kern-0.1em{\smaller A\kern-0.2em R}}}

\def\s     {\ensuremath{s}\xspace}

\def\pipi  {\ensuremath{\pi^+\pi^-}\xspace}

\def\Kbar  {\kern 0.2em\overline{\kern -0.2em K}{}\xspace}

\def\Kz    {\ensuremath{K^0}\xspace}
\def\Kzb   {\ensuremath{\Kbar^0}\xspace}
\def\KzKzb {\ensuremath{\Kz \kern -0.16em \Kzb}\xspace}
\def\Kp    {\ensuremath{K^+}\xspace}
\def\Km    {\ensuremath{K^-}\xspace}

\def\KpKm  {\ensuremath{\Kp \kern -0.16em \Km}\xspace}

\def\Dbar    {\kern 0.2em\overline{\kern -0.2em D}{}\xspace}

\def\Dz      {\ensuremath{D^0}\xspace}
\def\Dzb     {\ensuremath{\Dbar^0}\xspace}
\def\DzDzb   {\ensuremath{\Dz {\kern -0.16em \Dzb}}\xspace}
\def\Dp      {\ensuremath{D^+}\xspace}
\def\Dm      {\ensuremath{D^-}\xspace}

\def\DpDm    {\ensuremath{\Dp {\kern -0.16em \Dm}}\xspace}
\def\Dstar   {\ensuremath{D^*}\xspace}

\def\Dstarp  {\ensuremath{D^{*+}}\xspace}

\newcommand{\dsp}{\ensuremath{\Dstarp}\xspace}

\def\B       {\ensuremath{B}\xspace}
\def\Bbar    {\kern 0.18em\overline{\kern -0.18em B}{}\xspace}

\def\Bz      {\ensuremath{B^0}\xspace}
\def\Bzb     {\ensuremath{\Bbar^0}\xspace}
\def\BzBzb   {\ensuremath{\Bz {\kern -0.16em \Bzb}}\xspace}
\def\Bu      {\ensuremath{B^+}\xspace}
\def\Bub     {\ensuremath{B^-}\xspace}

\def\BpBm    {\ensuremath{\Bu {\kern -0.16em \Bub}}\xspace}

\def\BorBbar    {\kern 0.18em\optbar{\kern -0.18em B}{}\xspace}
\def\DorDbar    {\kern 0.18em\optbar{\kern -0.18em D}{}\xspace}
\def\KorKbar    {\kern 0.18em\optbar{\kern -0.18em K}{}\xspace}

\mathchardef\Upsilon="7107
\def\Y#1S{\ensuremath{\Upsilon{(#1S)}}\xspace}

\mathchardef\Deltares="7101
\mathchardef\Xi="7104
\mathchardef\Lambda="7103
\mathchardef\Sigma="7106
\mathchardef\Omega="710A

\def\Deltabar{\kern 0.25em\overline{\kern -0.25em \Deltares}{}\xspace}
\def\Lbar{\kern 0.2em\overline{\kern -0.2em\Lambda\kern 0.05em}\kern-0.05em{}\xspace}
\def\Sigbar{\kern 0.2em\overline{\kern -0.2em \Sigma}{}\xspace}
\def\Xibar{\kern 0.2em\overline{\kern -0.2em \Xi}{}\xspace}
\def\Obar{\kern 0.2em\overline{\kern -0.2em \Omega}{}\xspace}
\def\Nbar{\kern 0.2em\overline{\kern -0.2em N}{}\xspace}
\def\Xb{\kern 0.2em\overline{\kern -0.2em X}{}\xspace}

\newcommand{\tev}{\ensuremath{\mathrm{\,Te\kern -0.1em V}}\xspace}
\newcommand{\gev}{\ensuremath{\mathrm{\,Ge\kern -0.1em V}}\xspace}
\newcommand{\mev}{\ensuremath{\mathrm{\,Me\kern -0.1em V}}\xspace}
\newcommand{\kev}{\ensuremath{\mathrm{\,ke\kern -0.1em V}}\xspace}
\newcommand{\ev}{\ensuremath{\mathrm{\,e\kern -0.1em V}}\xspace}
\newcommand{\gevc}{\ensuremath{{\mathrm{\,Ge\kern -0.1em V\!/}c}}\xspace}
\newcommand{\mevc}{\ensuremath{{\mathrm{\,Me\kern -0.1em V\!/}c}}\xspace}
\newcommand{\gevcc}{\ensuremath{{\mathrm{\,Ge\kern -0.1em V\!/}c^2}}\xspace}
\newcommand{\mevcc}{\ensuremath{{\mathrm{\,Me\kern -0.1em V\!/}c^2}}\xspace}

\def\m    {\ensuremath{{\rm \,m}}\xspace}

\def\mus  {\ensuremath{\rm \,\mus}\xspace}

\def\mus        {\ensuremath{\,\mu{\rm s}}\xspace}    

\def\ra                 {\ensuremath{\rightarrow}\xspace}

\def\pep2{PEP-II}

\newcommand{\chisq}{\ensuremath{\chi^2}\xspace}

\def\gsim{{~\raise.15em\hbox{$>$}\kern-.85em
          \lower.35em\hbox{$\sim$}~}\xspace}
\def\lsim{{~\raise.15em\hbox{$<$}\kern-.85em
          \lower.35em\hbox{$\sim$}~}\xspace}

\xspace

\def\jetset74   {\mbox{\tt Jetset \hspace{-0.5em}7.\hspace{-0.2em}4}\xspace}

\long\def\inst#1{\par\nobreak\kern 4pt\nobreak
    {\it #1}\par\vskip 10pt plus 3pt minus 3pt}

\begin{document}
{\pagestyle{empty}


\begin{flushright}
\babar-CONF-\BABARPubYear/\BABARConfNumber \\
SLAC-PUB-\SLACPubNumber \\
hep-ex/0409047 \\
September 2004 \\
\end{flushright}

\par\vskip 3cm

\begin{center}
\Large \bf Measurement of $B \rar \Dst$ Form Factors 
in the Semileptonic Decay \Bbartodstarlnu
\end{center}
\bigskip

\begin{center}
\large The \babar\ Collaboration\\
\mbox{ }\\
\today
\end{center}
\bigskip \bigskip

\begin{center}
\large \bf Abstract
\end{center}

We present a preliminary measurement of $R_1$, $R_2$, and $\rho^2$, which
are the three parameters used to characterize the $B\rar D^*\ell\bar\nu_e$
form factors ($A_1$, $V$, and $A_2$). We use $85 \times 10^6$ $B\bar
B$-pairs accumulated on the $\Upsilon(4{\rm S})$ resonance at PEP-II. In
this analysis we use the decay mode $\bar B^0\rar D^{+*}e^-\bar\nu$ and
its charge conjugate. The $D^{*+}$ is reconstructed in the channel
$D^{*+}\rightarrow D^0\pi^+$ and the $D^0$ in the channel
$D^{0}\rightarrow K^{-}\pi^{+}$. We parameterize the form factors in terms
of their ratios (determined by the parameters $R_1$ and $R_2$) and the
common slope $\rho^2$ in the variable $w$ (a quantity related to the
momentum transfer in the decay process). These parameters are determined
via an unbinned maximum likelihood fit to the distributions in four
kinematic variables (three decay angles and $w$). The results are $R_1=
1.328 \pm 0.055\pm 0.025\pm 0.025$ and $R_2=0.920 \pm 0.044\pm 0.020\pm
0.013 $ for the ratios and $\rho^2= 0.769\pm 0.039 \pm 0.019\pm 0.032 $
for the slope. The stated errors are the statistical uncertainty from the
data, statistical uncertainty from the size of the Monte Carlo sample and
the systematic uncertainty, respectively.

\vfill
\begin{center}

Submitted to the 32$^{\rm nd}$ International Conference on High-Energy Physics, ICHEP 04,\\
16 August---22 August 2004, Beijing, China

\end{center}

\vspace{1.0cm}
\begin{center}
{\em Stanford Linear Accelerator Center, Stanford University, 
Stanford, CA 94309} \\ \vspace{0.1cm}\hrule\vspace{0.1cm}
Work supported in part by Department of Energy contract DE-AC03-76SF00515.
\end{center}

\newpage
} 

\begin{center}
\small

The \babar\ Collaboration,
\bigskip

%
B.~Aubert,
R.~Barate,
D.~Boutigny,
F.~Couderc,
J.-M.~Gaillard,
A.~Hicheur,
Y.~Karyotakis,
J.~P.~Lees,
V.~Tisserand,
A.~Zghiche
\inst{Laboratoire de Physique des Particules, F-74941 Annecy-le-Vieux, France }
% ...

\end{center}\newpage

\section{Introduction}
\label{sec:Introduction}
The decay $B\rar D^*\ell\nu$
can be described by three form
factors: two axial form factors, $A_1$ and $A_2$, and one vector form
factor, $V$.
They are functions of the momentum transfer $q^2$ (or equivalently,
$w$, defined below).  Measurements of these form factors provide us
with a ``laboratory'' for testing the predictions of heavy quark
effective theory (HQET)
\cite{NeubertPhysReport}.  These form factors are related to each
other by heavy quark symmetry (HQS) through the formalism of heavy
quark effective theory.  Deviations from the HQET relationships can be
computed as corrections to the theory. They can also in principle be
measured; the parameters we adopt for this analysis are inspired by
HQET, but allow for such deviations.

We introduce here a novel method of extracting these parameters
from the data. We use an unbinned-maximum-likelihood method, but
introduce approximations that allow us to correct efficiency and
resolution with the limited Monte Carlo (MC) data sample available.
In Sec. \ref{sec:Analysis} we sketch out how these approximations
work and how we evaluate their impact on the errors.

Improved measurements of the form factor parameters $R_1$, $R_2$ and
$\rhosq$ yield a significant reduction in the systematic error
obtainable on the Cabibbo-Kobayashi-Maskawa (CKM) matrix element
$V_{cb}$.  In this analysis, though we restrict ourselves to the
cleanest available decay channel, we still obtain nearly 17 times as
many reconstructed $B\rar D^*\ell\nu$ decay candidates as the
pioneering CLEO analysis~\cite{CLEO}. This already produces a
substantial improvement in the error achieved on the parameters $R_1$
and $R_2$ needed for extracting $V_{cb}$. Further, a better
understanding of this channel, which is the dominant background to $B
\ra X_u \ell \nu$, is also of great importance to the study of
inclusive and exclusive $B\ra X_u
\ell \nu$ decays and the determination of \vub.

With the addition of more data and the use of the other available
decay modes we will be able to further probe the consistency of HQET
predictions in the near future.

\section{Formalism}
\label{sec:Formalism}


\label{sec:wdescrip}

\begin{figure}[ht]
\begin{center}
{\parbox{6cm}
{\resizebox{!}{4cm}{\includegraphics{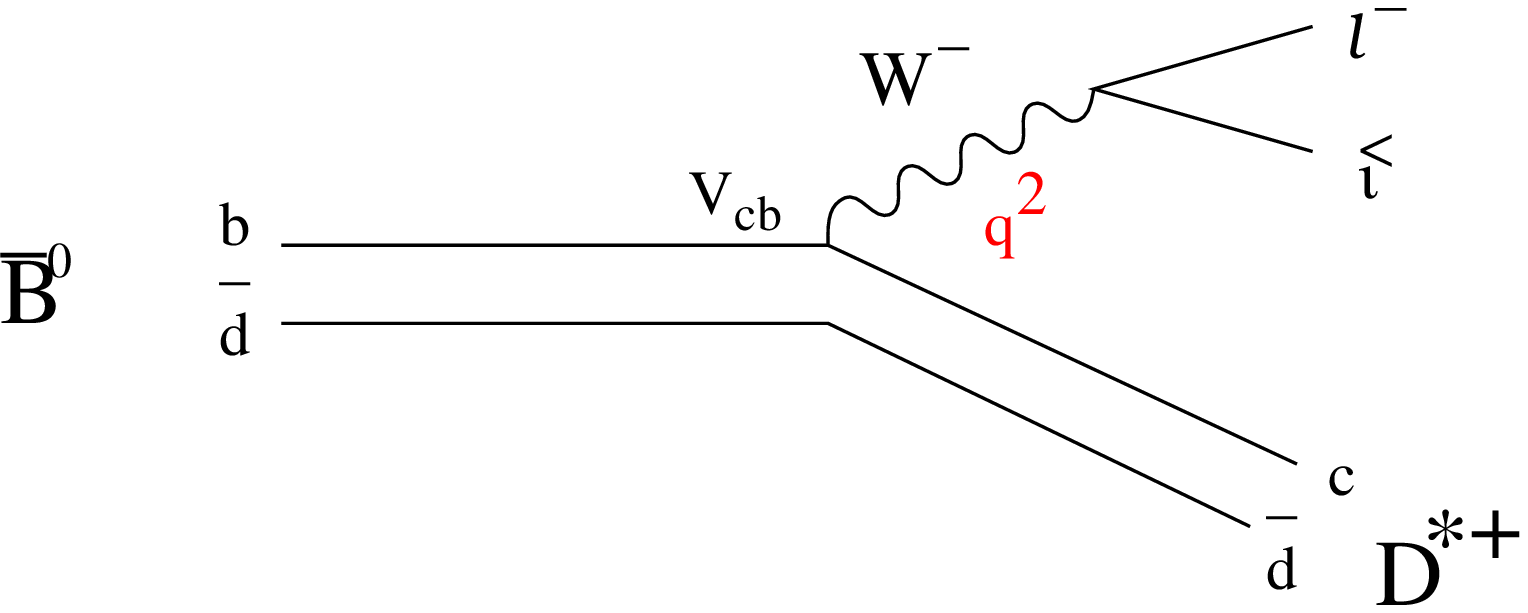}}
}}
\end{center}
\caption{Quark level diagram that leads to the decay  \bbartodstarlnu.}
\label{fig:quarkleveldiag}
\end{figure}

The Feynman diagram for the decay \Bbartodstarlnu\ is shown in
Fig.~\ref{fig:quarkleveldiag}.  

The square of momentum transfer from the $B$ to the $D^*$, $q^2$, is linearly
related to another Lorentz invariant ($w$) by

\be
\recow \equiv {\mb^2 + \mdstar^2 -q^2 \over 2 \mb \mdstar} =
v_B \cdot \vdstr = { p_B \cdot p_{D^*} \over \mb \mdstar}
\ee 
where $\mb$ and $\mdstar$ are the masses of the $B$ and the $D^*$
mesons, $p_B$ and $p_{D^*}$ are their four momenta, and $v_B$ and
$v_{D^*}$ are their four-velocities. In the $B$ rest frame $w$ reduces
to the Lorentz boost factor $\gamma_{D^*} = (E_{D^*}/M_{D^*}) $.

The range of $w$ and $q^2$ are restricted by the kinematics of the
decay with $q^2=0$ corresponding to
\be
w_{max}=\frac{\mb^2+\mdstar^2}{2\mb\mdstar}\approx1.504
\ee 
and $w_{min}=1$ corresponding to
\be
q_{max}^2=(\mb-\mdstar)^2\approx10.69 (\gevcc)^2.
\ee

\subsection{Kinematic Variables   }

\begin{figure}[h]
\begin{center}
{\parbox{10cm}
{\resizebox{!}{5cm}{\includegraphics{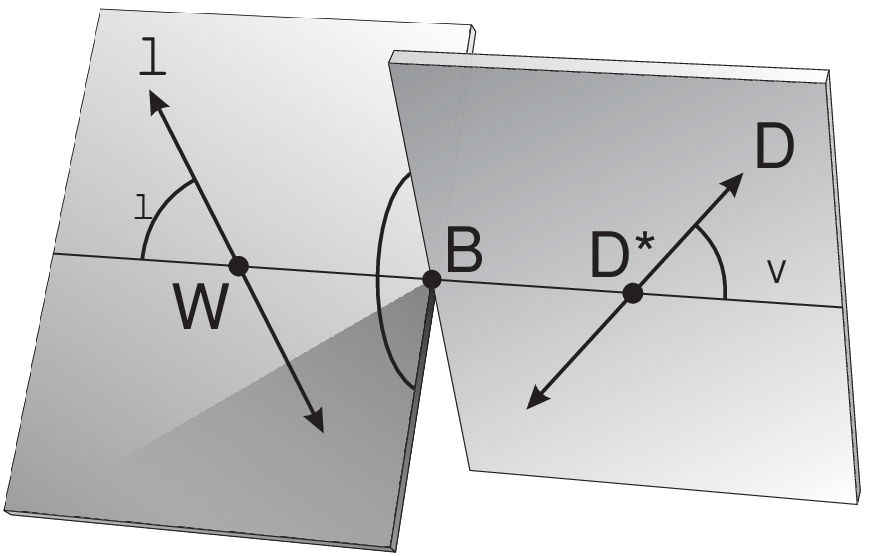}} }}
\end{center}
\caption{Kinematics of \Btodslnu\ decays.}
\label{fig:mesonleveldiag}
\end{figure}

  We choose to reconstruct the mode where the \Dstar\ decays to $D
  \pi$, and the $D$ to a $K \pi$ .  The outgoing particles are shown in
  Fig.~\ref{fig:mesonleveldiag}. The pion from the
  \Dstar\ decay and the electron are directly detected, and the $D$ is \recod\ from its
  daughters, the $K$ and $\pi$. 

\vspace{1cm}

In Fig. \ref{fig:mesonleveldiag} we define three angles, which,
along with \recow, constitute the four independent kinematic variables
we use to characterize this decay:

\bi 

\item{ $\ctl$,  the cosine of the angle between the direction of the lepton 
in the virtual $W$ rest frame, and the direction of the
virtual $W$ in the $B$ rest frame }

\item{ $\ctv$, the cosine of the angle between the direction of the 
$D$ in the $D^*$ rest frame, and the direction of the
$\dstar$ in the B rest frame }

\item{$\angchi$,  the azimuthal angle between the plane formed by the
$\dstar-D$ system and the plane formed by the $W-\ell$ system. }
\ei  

A \dstarlnu\ decay is completely characterized by the four kinematic
variables $w$, $\ctl$, $\ctv$ and $\chi$. Since the $B$ meson is
spinless, the direction of the $D^*-W$ axis 
relative to the $B$ direction is irrelevant to the dynamics.

\subsection{ Four Dimensional Decay Distribution}

\subsubsection{Helicity amplitudes}

The Lorentz structure of the $B\rar D^*\l\nu$ decay amplitude  can 
be expressed in terms of three amplitudes ($\ffH_+$, $\ffH_-$, and
$\ffH_0$), which
correspond to the three allowed polarization states of the
\Dstar\ (two transverse and one longitudinal). These amplitudes can be 
completely specified in terms of the axial and 
vector \FF s as follows\cite{NeubertPhysReport}:

\bea
\label{eq:hel_ampp1}
\ffH_{+}(w) \equiv -(\mb+M_{D^*}){A_{1}(w)}+ 2{ \momdstr \mb
\over \mb + M_{D^*}}{V(w)},
\\
\label{eq:hel_ampp2}
\ffH_{-}(w) \equiv -(\mb+M_{D^*}){A_{1}(w)}-
      2{ \momdstr \mb \over \mb + M_{D^*}}{V(w)},
\\
\label{eq:hel_ampp3}
\ffH_{0}(w)  \equiv  -{1 \over 2 M_{D^*} \sqrt{q^{2}}} \lbracket {A_{1}(w)}
        (\mb + M_{D^*})(\mb^{2}-M_{D^*}^{2}-q^{2})- \right. \\  \nonumber
\left. 4{\mb^2 \momdstr^2 \over \mb + M_{D^*} }{A_{2}(w)} ) \rbracket ,
\eea
where $\wp_D^*=M_{D^*}\sqrt{w^2-1}$ is the magnitude of the momentum of the $D^*$ in the $B$ rest frame.

 By contracting the relevant lepton and hadron tensors, neglecting
small terms of order $m^2_\ell/q^2$, and doing the
 phase space integrations, the full differential decay rate

\bea
\label{eq:dsdo}
\label{eq:thlpdf}
{d\Gamma (B\rar \dstarlnu) \over dw \; d\ctl \; d\ctv \; d\chi}  & = & 
{6G^{2}_{F}\vert V_{cb} \vert^{2} \mb\mdstar^2 r\sqrt{(w^2-1)} (1-2wr+r^2)  \over 8(4\pi)^{4}} \times  \nonumber \\
 & &\lb {H_{+}}^2(1-\ctl)^{2}\sin^{2}\theta_{V}+
   {H_{-}}^2(1+\ctl)^{2}\sin^{2}\theta_{V} \right.    \nonumber \\
 & & +4{H_{0}}^{2}\sin^{2}\theta_{\ell} \cos^{2}\theta_{V}  
       -2{H_{+}H_{-}}\sin^{2}\theta_{\ell} \sin^{2}\theta_{V}\cos(2\chi)  \nonumber  \\
 & & -4{H_{+}H_{0}}\stl(1-\ctl)\stv\ctv\cos\chi   \\
 & & \left. +4{H_{-}H_{0}}\stl(1+\ctl)\stv\ctv\cos\chi \rb  \nonumber
\eea
can be obtained in terms of the helicity amplitudes.  Here $r\equiv
{\lp M_{D^*} \over \mb \rp }$ is the ratio of the $D^*$ to the $B$
mass.  Details of the derivation can be found in
Neubert\cite{NeubertPhysReport}. The four-dimensional distribution of
$w$, $\ctl$ $\ctv$ and $\chi$ that is described by
Eq.(\ref{eq:thlpdf}) is the physical observable from which we can
extract the form factors.

\subsubsection{ Heavy Quark Symmetry Relationships  }
\label{sec:hqsrships}

HQS relates the three form factors to each other, as follows:

\be
\label{eq:a2}
A_2(w)=\frac{R_2(w)}{R^{*2}}\frac{2}{w+1}A_1(w),
\ee
\be
\label{eq:v}
V(w)=\frac{R_1(w)}{R^{*2}}\frac{2}{w+1 } A_1(w),
\ee
where we have defined the constant
\be
R^*\equiv {   {2\sqrt{M_{B} M_{\Dstar}} }  \over {(M_B+M_{\Dstar})}   }.
\ee
$R_1(w)=R_2(w)=1.0$, i.e., perfect HQS, implies that the form factors,
$A_2$ and $V$ 
are identical for all values of $w$.
Since HQS is not exact, $R_1$ and
 $R_2$ can differ from $1.0$ and exhibit some $w$-dependence.

$A_1$ can be related to the Isgur-Wise function  by

\be
A_1(w)=R^*\frac{w+1}{2}h_{A_1}(w),
\ee
where in the HQET limit $h_{A_1}$ is the Isgur-Wise function $\xi(w)$ \cite{isgurwise}. HQET
predicts $h_{A_1}(1)=\xi(1)=1.0$.

It is convenient to re-express the helicity
amplitudes $H_i$ ($i=\pm,0$) in terms of the functions  $R_1(w)$ and
$R_2(w)$  as follows


\bea
\mathit{\tilde{\ffH}_+} & = \lbrackt {\displaystyle \frac {\sqrt{(1 - 2\,\recow\,r + r^{2})}\,\lp 1 - 
\sqrt{{\displaystyle \frac {\recow - 1}{\recow + 1}} }\,\mathit{ \rone(w) }\rp}{(
1 - r)}} \rbrackt , \nonumber \\
\mathit{\tilde{\ffH}_0}  &  =    \lb 1 + {\displaystyle \frac {(\recow - 1)\, (1 - \mathit{
\rtwo(w) })}{1 - r}} \rb , \\
\vsps
\mathit{\tilde{\ffH}_-} & =  \lbrackt {\displaystyle \frac {\sqrt{(1 - 2\,\recow\,r + r^{2})}\,\lp 1 + 
\sqrt{{\displaystyle \frac {\recow - 1}{\recow + 1}} }\,\mathit{\rone(w) }\rp}{(
1 - r)}} \rbrackt ,\nonumber \\
\label{eq:helampsFunctsOfparams} \nonumber 
\eea
where the helicity amplitudes that appear in
Eqs.(\ref{eq:hel_ampp1}-\ref{eq:hel_ampp3}) differ from these only by
a common kinematic factor times $\haone$:
\bea
  {\ffH}_i =  -\mb\frac{R^*(1-r^2)(w+1)}{2\sqrt{1-2wr+r^2}}h_{\ffA_1}(w)\tilde{H}_i  \\
\label{eq:helampDefs} \nonumber. 
\eea
The function $h_{A_1}(w)$ can be conveniently parameterized by exploiting 
the fact that $w-1$ is a small parameter ($ \lsim 0.5$)
to express it as a power series expansion

\be
h_{A_1}(w)=h_{A_1}(1)\left(1-\rho^2(w-1)+c(w-1)^2+...\right)
\ee
where $\rho^2$ is called the slope and $c$ is called the curvature
(note that in the literature $\rho^2$ is often referred to as
$\rhosqaone$).  Corrections to HQET
modify $ h_{A_1}(1) $ and thus lead to deviations from the HQET limit
of $h_{A_1}=1.0$.  However, in this analysis we only deal with the
shape and relative normalization of the form factors, and consequently
the overall normalization is irrelevant.

For this preliminary analysis we only use the expansion to first order in $(w-1)$, so
we set $c=0$.  In our baseline analysis we treat $R_1$ and
$R_2$ as constant.  However, we also show results in which we extend
the analysis to explore the impact of $w$-dependence.

\subsection{ \FFcap\ Predictions }

As discussed in Sec. \ref{sec:hqsrships}, for infinitely massive $b$ and $c$ quarks, we expect
$\rone=\rtwo=1.0$, but for finite masses they are modified by both perturbative $\as$
and non-perturbative $\lqcdomx$ corrections.

Calculating higher-order loop corrections to the \FF s yields
expansions of the form:

\bea
 {\rone}(w) & = & 1 + \lb \as (...) + \asq  (...) \rb + \lqcdomx (...) , \\ 
 {\rtwo }(w) & = & 1  + \lb \as (...) + \asq  (...)  \rb + \lqcdomx (...).
\label{eq:correcs}
\eea

The coefficients of the $\as$ terms (shown as ellipses) have been
calculated perturbatively up to second order, which gives confidence
that they are accurate to about one percent (see
\cite{LigetiGrinstein}).  The coefficients of the $\lqcdomx$ terms are combinations
of quantities called ``subleading Isgur-Wise functions.''

Different models in the HQET framework evaluate the subleading Isgur-Wise function correction
terms differently, resulting in a variety of predictions for $\rone(w)$
and $\rtwo(w)$.  Neubert \cite{NeubertPhysReport}, based on work with
collaborators, in the early 90's predicted

\bea
  \rone(\recow) & = & 1.35-0.22(w-1)+0.09(w-1)^2,  \label{eq:neubr1r2} \\
 \rtwo(\recow) & = & 0.79+0.15(w-1)-0.04(w-1)^2.
\eea

More recently, Caprini {\it et.al.} \cite{Caprini} using spectral functions, dispersion relations,
and heavy quark symmetry to evaluate the non-perturbative terms
predict

\bea
  {\rone}( w) & = & 1.27-0.12( w-1)+0.05( w-1)^2,\\
 {\rtwo }( w) & = & 0.80+0.11( w-1)-0.06( w-1)^2.
\label{eq:caprini}
\eea
Ligeti and Grinstein \cite{LigetiGrinstein} using similar HQET machinery find

\bea
  {\rone}( w) & = & 1.25-0.10( w-1), \label{eq:ligeti} \\
 {\rtwo }( w) & = & 0.81+0.09( w-1).
\eea

Whereas all the above HQET-based predictions predict \FF\ values
within a fairly narrow range, older predictions relying on {\it ad-hoc}
potential models vary widely, {\it e.g.}, Close \& Wambach using a simple
quark model predict \cite{CloseWambach}

\bea
  {\rone }( w) & = & 1.15-0.07( w-1), \label{eq:cwr1r2} \\
  {\rtwo }( w) & = & 0.91+0.04( w-1). 
\eea

It can be seen that in all the above predictions the \coef s of the $(w-1)$
and $(w-1)^2$ terms are quite small because \rone\ and \rtwo\ are by
construction as ratios expected to vary only slightly with $w$, where
$\haone$ has no such restriction.

\section{The \babar\ Detector and Dataset}
\label{sec:babar}


The data set used in this analysis was collected with the \babar\
detector at the \pep2\ storage ring during the period between 2000 and
2002. It corresponds to $84 \ifb$ collected on the \upsfs\ resonance,
which yields approximately $85\times10^6$ $B\bar B$-pairs. There are
$\approx8.5\times10^6$ $\btodstarlnu$ decays in this sample of which
we have reconstructed 16,386 $\bar B^0\rar D^{*+}e^-\bar\nu_e$ (or
charge conjugate) candidates using only the $D^0\pi^+$ decay of the
$D^*$ and the $K^-\pi^+$ decay mode of the $D^0$.

The \babar\ detector is described elsewhere in
detail~\cite{ref:babar}.  This analysis uses four of the five subdetectors of
\babar: the silicon vertex tracker, the drift chamber, a Cerenkov-light-based 
particle identification detector, and the electromagnetic calorimeter. This analysis depends critically on the silicon vertex
tracker to reconstruct the low momentum pions produced by the decay
$D^{*+}\rightarrow D^0\pi^+$, about two-thirds of which do not
traverse more than about a fourth of the drift chamber.

\section{Reconstruction and Event Selection}
\label{sec:recoAndSel}
We reconstruct the lepton, and the $D^ *$ through its decay products.
After \recon\ of all tracks, we choose cuts that select $B\rar
D^*\ell\nu$ decays and determine the kinematic variables
$w$, $\ctl$, $\ctv$ and $\chi$. 

\subsection{Event selection}

In this analysis we only consider the decay channel
$D^{*+}e^-\bar\nu_e$ with the $D^{*+}$ decaying to $D^0\pi^+$ and the
$D^0$ decaying in the $K^-\pi^+$ mode. In case of multiple $D^{*}e\nu$
candidates in a given event we choose the candidate with $K\pi$ mass
($m_{K\pi}$) closest to the $D^0$ mass.

For event selection we use the procedure developed for our $V_{cb}$
analysis~\cite{babarVcb}, and we also apply the same selection
criteria.  We summarize the most salient of these cuts here:

\begin{itemize}

\item{
The momentum of the lepton in the center-of-mass (C.M.) frame $p^*_\ell$
is required to be larger than $1.2 \gevc$. This criterion selects $B$
semi-leptonic decays and suppresses continuum ($c\bar c$) and cascade
($b\rar c\rar \ell$) backgrounds.  }

\item{
The momentum of the $D^*$ in the C.M. frame must be between $0.5$ 
and
$2.5 \gevc$.  }

\item{
The soft pion from the $D^*$ decay must have a transverse momentum
greater than $50 \mevc$.  }

\item{
The $\chi^2$ probability of the fit of the $D^*\ell$ vertex to the
beamspot constraint must be greater than $1\%$.
}

\item {To further suppress continuum background, 
we select only candidates with $|\cos\theta_{\rm thrust}|<0.85$, where
$\theta_{\rm thrust}$ is the angle between the thrust axis of the
$D^*\ell\nu$ candidate and the thrust axis of the rest of the event.
}

\end{itemize}

The cosine of angle $\theta_{BY}$ between the direction of the $B$ and the
direction of the \dstarell\ system can be computed from the kinematics
of the decay (see next Section).  Candidates with $\cosby$ between
$-10$ and $+6$ have been used to estimate background.  We include only
events that have $|\cosby| \leq 1.2$ in the final sample.

The final selection is based on $\deltam=m_{K\pi\pi_s}-m_{K\pi}$. For
candidates in which the $\pi_s$ track is reconstructed with twelve or
more drift chamber hits we cut on the mass difference
$0.144<\deltam<0.147$ \gevc. For the less well-measured case with fewer than
twelve drift chamber hits we loosen the cut to
$0.143<\deltam<0.148$ \gevc. About three-quarters of the candidates are in
the latter category.

\begin{figure}[htp]
\begin{center}
{\parbox{14cm}
  {\resizebox{!}{14cm}{\includegraphics{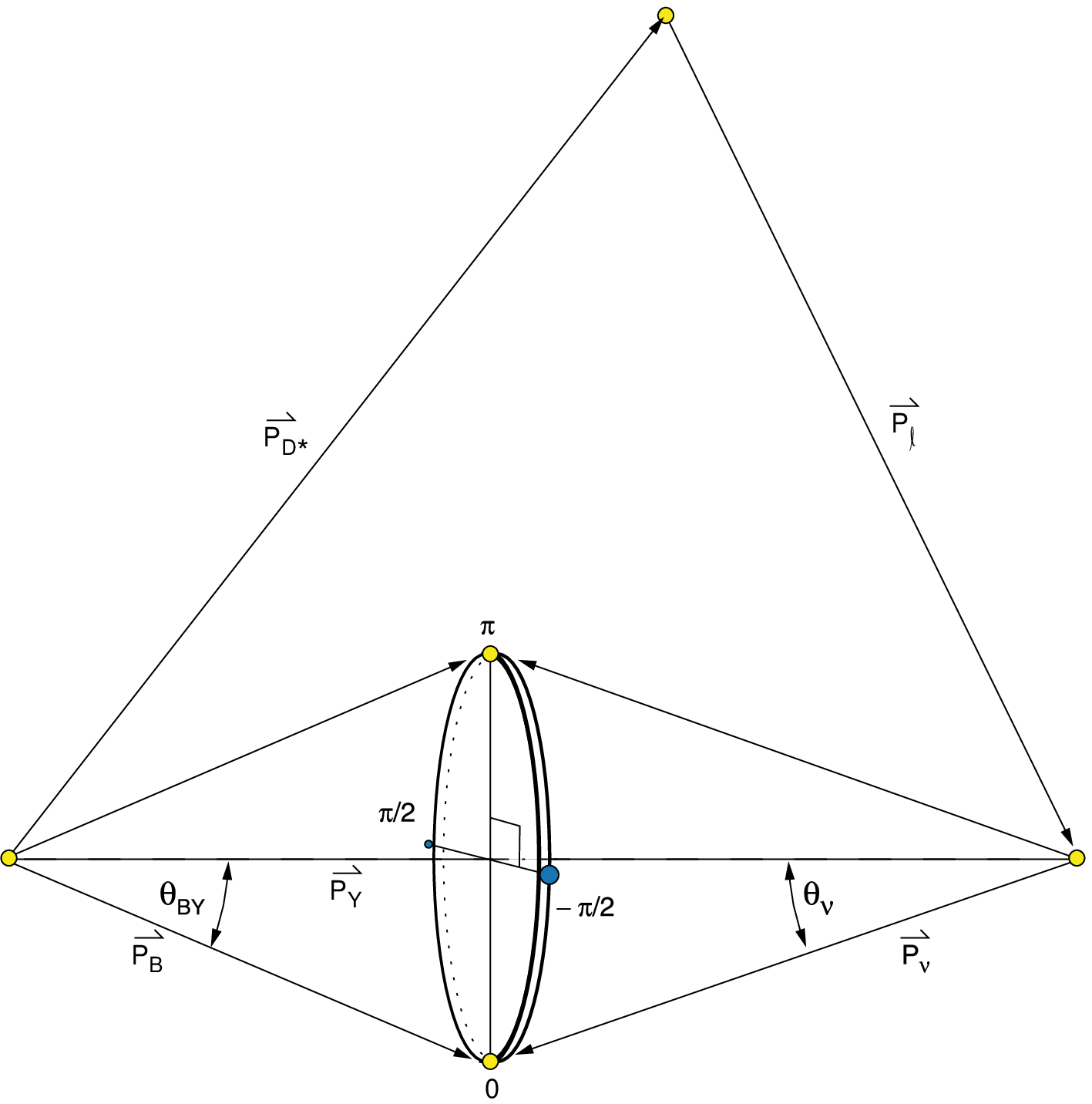}}
}}
\end{center}
\caption{
Diagram of $\cos\theta_{BY}$ reconstruction. The points at $\phi_{BY}=0$ and $\pi$ are in the \dstarell\ plane. The
points at $\pm \pi/2$ are out of the plane.
}
\label{fig:cosbyfig}
\end{figure}

\subsection{Kinematic variable determination}

Without the neutrino we do not have sufficient information to
fully reconstruct the kinematic variables $w$, $\ctl$,
$\cos\theta_V$ and $\chi$. However, we can use the measured variables
to construct an adequate approximation. Using energy-momentum
conservation and assuming that the neutrino mass $m_\nu$ is zero, we
have

\be
\label{eq:mnu}
0=m^2_\nu=M_B^2+M_Y^2-2E_BE_Y+2\wp_B \wp_Y \cos\theta_{BY},
\ee
where $p_Y=p_D^*+p_\ell$ is the four momentum of the combined $D^*$
and lepton, $M_Y^2=p_Y^2$ is the mass squared and $\wp_Y$ is the three-momentum.
The $B$ meson energy $E_B$
and three-momentum $\wp_B$ can be estimated from the energies of the
colliding beam  particles, so we can solve for $\cos\theta_{BY}$ as
follows

\be
\label{eq:cosby}
\cos\theta_{BY}=-\frac{M_B^2+M_Y^2-2E_BE_Y}{2 \wp_B \wp_Y }.
\ee

Thus we can determine the angle between the $B$ and the direction
($\hat Y=\vec p_Y/\wp_Y$) of the \dstarell\ system, but we do not know
its azimuthal angle $\phi_{BY}$ around this direction. This is
illustrated in Fig.~\ref{fig:cosbyfig}. The direction of the $B$
must lie on the cone around $\hat Y$ defined by the angle
$\theta_{BY}$.

For each possible $\phi_{BY}$ we can compute the kinematic variables
$w$, $\ctl$, $\ctv$, $\chi$. Since we do not know which direction is
correct we perform an average to estimate each variable.  We average
over four points: two in the $D^*$-lepton plane corresponding to the
azimuthal angles $\phi_{BY}=0$ and $\pi$ and two points out of the
plane corresponding to the angles $\pm\pi/2$. Further, since $B\bar B$
production follows a $\sin^2\theta_B$ distribution in the angle between
the $B$ direction and the beam collision axis, we weight the \kv s
evaluated at each point by the value $\sin^2\theta_B$ corresponding to
the $B$ direction for that point.

\begin{figure}[hbp]
\begin{center}
{\parbox{12cm}
  {\resizebox{!}{12cm}{\includegraphics{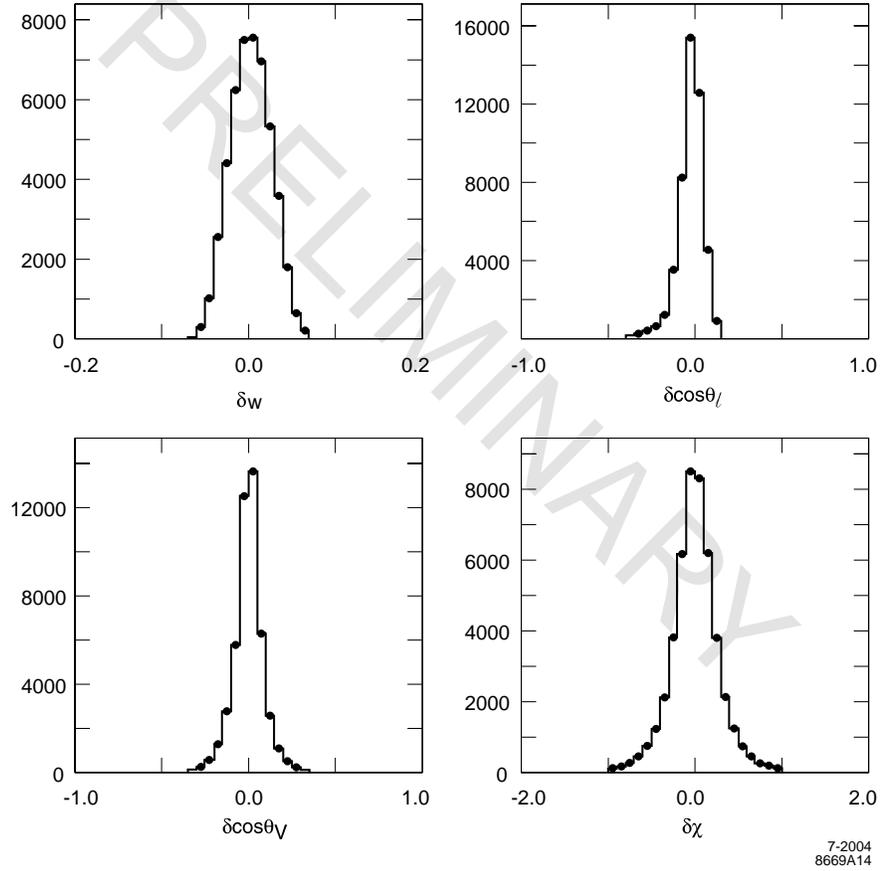}}
}}
\end{center}
\caption{
MC assessment of the experimental resolution for the variables $w$,
$\ctl$, $\ctv$ and $\chi$. For each variable we plot the difference
beween reconstructed and generated values.  The \resln\ is generally
small compared to the ranges of the variables as shown
in Fig.~\ref{fig:kinvar-plot}.}
\label{fig:kinvarres}
\end{figure}

Fig.~\ref{fig:kinvarres} illustrates the resolution achieved by this
`partial reconstruction' technique.  It is seen that the core widths
for each \resln\ \dist\ are relatively small compared to the full
width of each \kv.  The resolution is dominated by the average over
the $B$ direction; detector resolution makes only a relatively minor
contribution. The low-side tail on $\ctl$ can be attributed to final
state radiation.

The resolutions of the four kinematic variables are highly
correlated. A simple factorized resolution function in which the resolution 
is represented by a product of independent function for each variable
fails to capture these correrlations. Thus, we are dependent on the 
Monte Carlo simulation to account for resolution effects.

The distributions of the reconstructed kinematic variables $w$,
$\ctl$, $\ctv$ and $\chi$ are displayed in
Fig.~\ref{fig:kinvar-plot}. The shaded region is the distribution of
the background as estimated from the MC simulation using the method
described in Sec. \ref{sec:Analysis} below.  It can be observed
that the \bkgd\ \dist\ is much smaller than the signal contribution.

\begin{figure}[ht]
\begin{center}
{\parbox{12cm}
  {\resizebox{!}{12cm}{\includegraphics{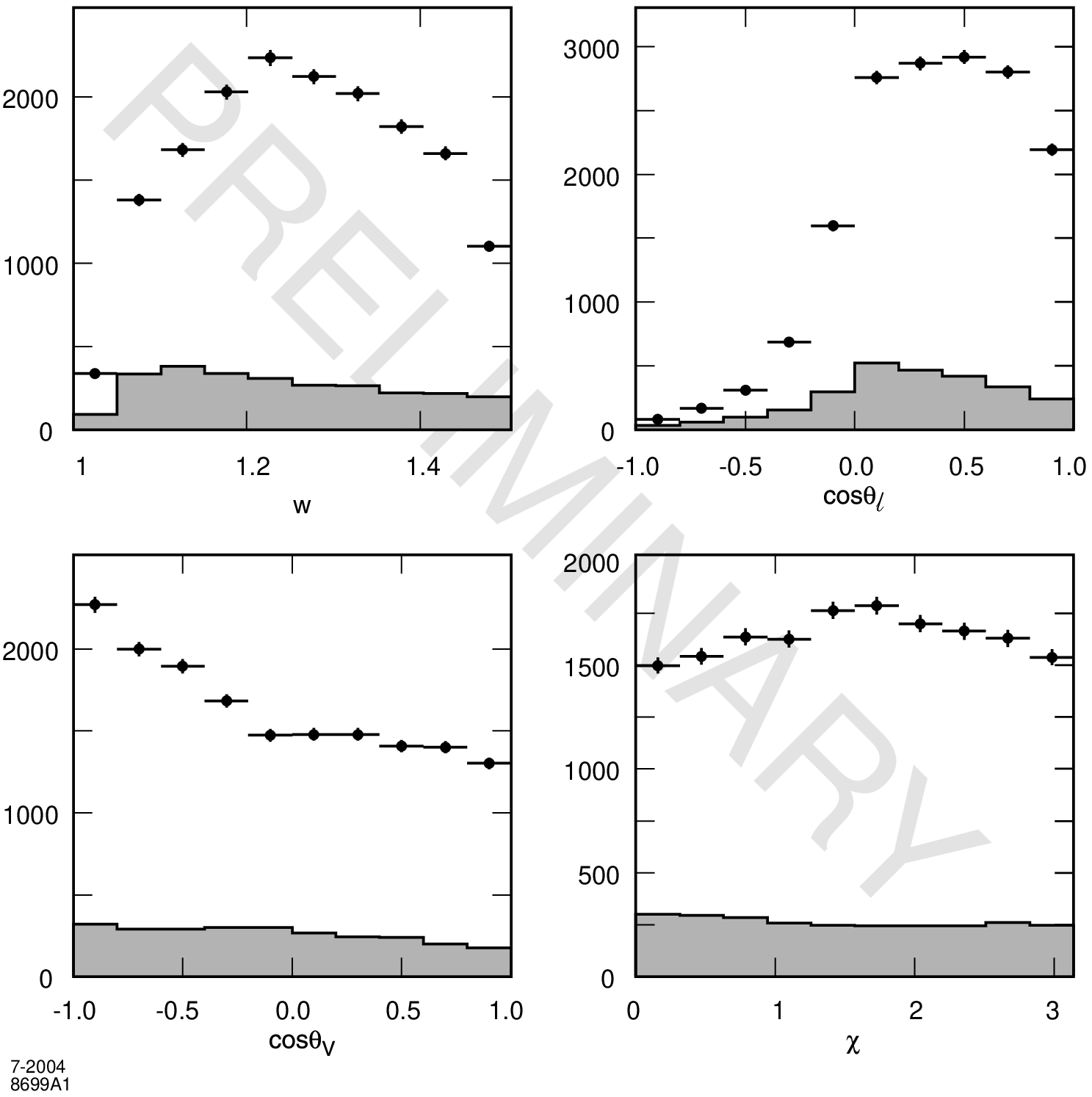}}
}}
\end{center}
\parbox{0.85\textwidth}{
\caption{\label{fig:kinvar-plot}
Distribution of kinematic variables $w$, $\ctl$, $\ctv$ and $\chi$ for
selected events with estimated backgrounds (shaded region).  }
}

\end{figure}

\label{sec:mc}
\section{Simulation}
This analysis is dependent on Monte Carlo (MC) simulation to model the
efficiency and the background distributions. 
The degree to which we can trust our simulation to model both the detector and 
the underlying physics processes  largely determines our
systematic errors.

For the detector simulation we use \babar's GEANT4 based
simulation~\cite{babarsim}. The simulation has been validated and
sophisticated correction procedures have been developed using many control
samples.  For simplicity, we do not apply these correction factors,
but use them to evaluate systematic errors (which turn out to be small
for this analysis, thereby justifying using them for {\it a
posteriori} error estimation rather than as correction factors). Event
generation and particle decay are modeled using the package
EvtGen~\cite{evtgen}.

\begin{figure}[htp]
\begin{center}
{\parbox{14cm}
  {\resizebox{!}{14cm}{\includegraphics{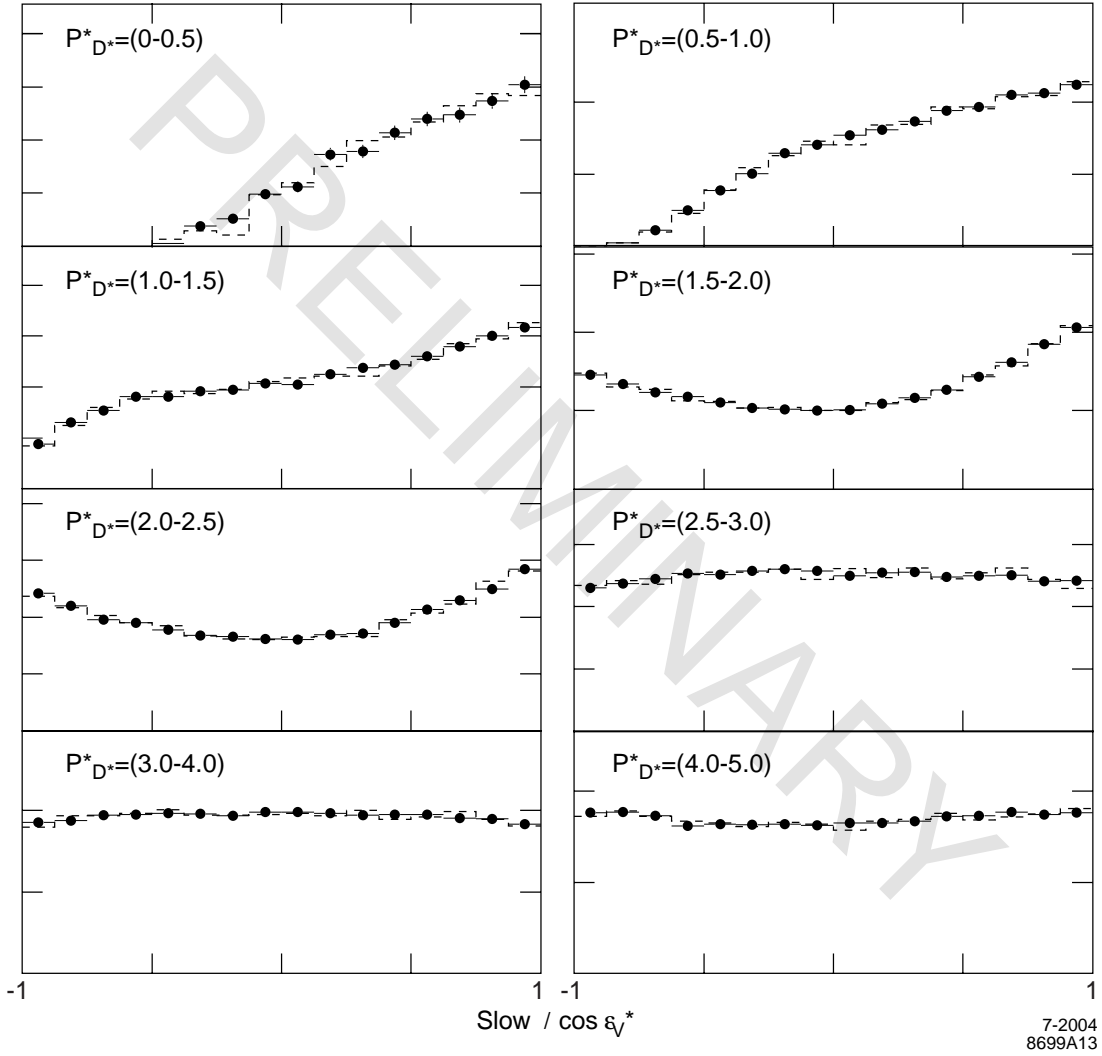}}
}}
\end{center}
\caption{ \label{fig:dsthel}
Distributions of the cosine of the helicity angle $\cos\theta_{V}$ for 
data (points) and MC (histogram) 
in bins of $D^*$ center-of-mass
momentum. The momentum range for each plot is given its upper left-hand corner. 
The MC has been fit to the data as described in the text.
}
\end{figure}

\ssec{Slow pions}
Of particular importance to this analysis is the modeling of the
efficiency for detecting low-momentum pions. This is a difficult task
since low momentum pions are lost through the interplay of acceptance,
decay-in-flight, and stopping and scattering 
in the beam pipe or vertex detector.  The
performance of our simulation can be seen in Fig.~\ref{fig:dsthel},
which shows the distribution of cosine of the helicity angle
($\cos\theta_V$) for $D^{*+}\rightarrow D^0\pi^+$ obtained from
inclusively produced $D^*$ mesons for both the data and the MC sample in bins
of center-of-mass momentum. The MC sample has been corrected by a factor of
the form
\be
\label{eq:fcorr}
f_{corr}=N\left(1+\alpha \cos\theta_V + \delta \cos^2\theta_V)\right)
\ee
where the normalization and the parameters $\alpha$ and $\delta$ have
been obtained from a fit to the data.  The three lowest momentum bins
are most relevant to $D^*\ell\nu$.  The excellent agreement between
the corrected MC and the data can clearly be observed in this figure.

The quadratic terms ($\delta$) may arise from differences in
$D^*$ polarization; they are large ($\sim 0.5$ and $0.7$) 
in bins 2 and 3 where we expect the $D^*$ spectrum to be dominated
by highly polarized few-body $B$ meson decays.
The linear terms can only be attributed to differences in
efficiency.  The linear terms for the first three bins are small; i.e., they are $0.037\pm 0.16$, $-0.023\pm 0.024$ 
and $-0.016 \pm 0.009$.

\ssec{Final state radiation}
The program PHOTOS~\cite{Was} is used to model the effects of final
state radiation (FSR). PHOTOS uses QED to second order in
$\alpha_{em}$ (up to two FSR photons can be produced) and is known to
provide an accurate simulation of the FSR effects.

\ssec{Signal}
To simulate the signal we use Eq.~(\ref{eq:dsdo}) for the distribution
of the decay products.  The MC samples are generated with the default
parameters $R_1=1.180$, $R_2=0.720$ and $\rhosq=0.920$ \cite{Ryd}.

\ssec{Other semileptonic decays}
A major source of background is other semileptonic $B$ decays. Aside from 
branching fractions for 
the decay modes $D\ell\nu$ and the
signal mode $D^*\ell\nu$, only that for the
mode $B \rar D_1\ell\nu$ has been measured~\cite{pdg2002}.  Thus, the 
other
branching fractions used in the MC simulation are based on theory. Until such
time as we can measure them, we must accept quite large possible
variations in their numerical input values in our simulation.

\section{Analysis Method}
\label{sec:Analysis}
\subsection{Fitting}
\label{sec:fitting}

Our basic approach to extracting the $D^*\ell\nu$ form factors is to
perform an unbinned maximum likelihood fit to the full four-dimensional
(4D) distribution (usually referred to as the probability density
function or PDF in the context of likelihood fitting) specified by
Eq.~(\ref{eq:dsdo}). We parameterize the form factors in terms of the
HQET inspired parameters $R_1$, $R_2$ and $\rhosq$ as described in
Sec. \ref{sec:Formalism}.  Since theoretical predictions differ on
the $w$ dependence of $R_1$ and $R_2$ (see
Eqs.(\ref{eq:neubr1r2}-\ref{eq:cwr1r2}) ), we first give as our
baseline result the values obtained by treating them as \param s which
are constants over the entire range of $w$ ({\it i.e.} setting the
coefficients of the $(w-1)$ and $(w-1)^2$ terms to zero and fitting only for the
constant terms). This simplifies comparisons between
experiments. However, we also show how the results vary when using the
$w$-dependencies suggested by the full predictions of
Eqs.~(\ref{eq:neubr1r2}-\ref{eq:cwr1r2}).

In addition to assuming the form of the theoretical PDF, we must
account for the effects of resolution and efficiency on the kinematic
variable distribution. As our Monte Carlo samples are limited in size
and do not allow us to adequately map the resolution and efficiency as
a function of $w$, $\ctl$, $\ctv$ and $\chi$, we resort to the
approximations derived below to allow us to correct for the effect of 
efficiency and
resolution on the fitted parameters
using only  integrals of this efficiency. We must also account
for the residual background.

To account for the efficiency and resolution effects we adapt the approach 
first employed in our angular analysis of the 
decay $B\rightarrow J/\psi K^*$\cite{psikstar} (for more details see also~\cite{Gill}).
The PDF including resolution and
efficiency ($\tilde F$) is given by

\be
\label{eq:respdf}
\tilde F(\tilde x;\mu)=\int{dx \, \varepsilon(x) \, G(\tilde x; x) \, F(x;\mu)}
\ee
where $x$ represents the true variables ($w$, $\ctl$, $\ctv$ and
$\chi$), $\tilde x$ are the observed values of the variables and
$\mu$ represents the parameters ($R_1$, $R_2$ and $\rhosq$) that
determine the form factors.  The efficiency $\varepsilon(x)$ is the
fraction of events with parameters $x$ that will be -- on average --
detected and $G(\tilde x; x)$ is the probability that an event with true
parameters $x$ will be reconstructed with parameters $\tilde x$.

The log of the likelihood $L$ that we need to maximize is given by

\be
\label{eq:likelihood}
\ln L=\sum_i{\ln\left( \frac{\tilde F(\tilde x_i;\mu)}{N(\mu)}\right)}=\sum{ \ln \tilde F(\tilde x_i;\mu)}-N_{\rm signal}\times \ln N(\mu)
\ee
where the integral $N(\mu)\equiv\int{d\tilde x\tilde
F(\tilde x;\mu)}$ is needed to properly normalize the likelihood in
the presence of imperfect acceptance.  The sum is over our
data event sample. $N_{\rm signal}$ is the number of signal events in the data sample.

Now consider the following trivial modification (multiplication by
$1=F(x;\mu_{\rm mc})/F(x;\mu_{\rm mc}))$, where $\mu_{\rm mc}$ is the parameter set used to 
generate our Monte Carlo sample) of Eq.~(\ref{eq:respdf})

\be
\label{eq:respdfmod}
\tilde F(\tilde x;\mu)=\int{dx \, \varepsilon(x) \, G(\tilde x; x) \, F(x;\mu_{\rm mc})\times \frac{F(x;\mu)}{F(x;\mu_{\rm mc})}}
\ee

If $\mu$ is not too different from $\mu_{\rm mc}$ 
then the ratio $F(x;\mu)/F(x;\mu_{\rm mc})$ will not vary much
across the range where the resolution function is much different from
zero. In this case we can use $\tilde x$ to approximate $x$, which
allows us to pull this ratio out of the integral to obtain

\be
\label{pdfapprx}
\widetilde F(\tilde x;\mu)\approx\frac{F(\tilde x;\mu)}{F(\tilde x;\mu_{\rm mc})}\int{dx \, \varepsilon(x)G(\tilde x;x)F(x;\mu_{\rm mc})}
=F(\tilde x;\mu)\frac{\tilde F(\tilde x;\mu_{\rm mc})}{ F(\tilde x;\mu_{\rm mc})}.
\ee

When substituted into the expression for the log-likelihood this gives

\be
\label{eq:llapprox}
\ln L=\sum \ln F(\tilde x_i;\mu)-\sum \ln F(\tilde x_i;\mu_{\rm mc})+\sum \ln \tilde F(\tilde x_i;\mu_{\rm mc})-N_{\rm signal}\ln N(\mu).
\ee

Since terms that are independent of the fit parameters (constant
terms) do not affect the point at which the maximum will be
found, all the sums that depend on $\mu_{\rm mc}$ can be dropped. The
$\mu$-dependent piece has been factored from these constant terms. We
are left with a likelihood that depends only on the theoretical PDF
$F$ and on the integral over the resolution and efficiency
functions. Thus, there is no need for a detailed understanding of the
efficiency and resolution as a function of the kinematic
variables. All we need is a method for evaluating of the normalization integral
$N(\mu)$

We can use the technique of Monte Carlo integration to
evaluate the integral $N(\mu)$. We have

\be
\label{apprxintegral}
N(\mu)\approx\int{d\tilde x\tilde F(\tilde x;\mu_{\rm mc})\times\frac{F(\tilde x;\mu)}{F(\tilde x;\mu_{\rm mc})}}
	\approx\frac{1}{N_{\rm gen}}\sum_i{\frac{F(\tilde x_i;\mu)}{F(\tilde x_i;\mu_{\rm mc})}}
\ee
where we have used the fact that the Monte Carlo generates events
proportional to $\tilde F(\tilde x;\mu_{\rm mc})$ to approximate the
integral by a sum over MC events divided by the number of
events generated, $N_{\rm gen}$.

We call this technique the `resolution-efficiency correction'
(REC) method. It can be shown to account for both the biases produced by
resolution and efficiency if the Monte Carlo sample used was
generated with parameters sufficiently near the final fitted
values. As will be seen in the next section the small residual bias is
also correctable.

The last remaining step is to handle the background. Ordinarily, we
would add a PDF $B(\tilde x)$ that models the background. The PDF $\tilde F$ would be replaced with
\be
f\tilde F(\tilde x;\mu)+(1-f)B(\tilde x)
\ee
where $f$ is the signal fraction.  However, since we do not have a form
for the background before acceptance,
we can not achieve the
factorization of the parameter dependence from the efficiency and
resolution functions that leads to Eq. (\ref{eq:llapprox}).
To
avoid this problem we subtract the background directly in our
likelihood sum rather than adding it to our PDF. We replace our
log-likelihood function with the following `pseudo-likelihood'

\be
\label{pseudoll}
\ln \Lambda=\sum_i \ln F(x^{(i)}_{\rm data};\mu)-\sum_i w^{(i)}_{\rm bkgd}\ln F(x^{(i)}_{\rm bkgd};\mu)-N_{\rm signal}\ln N(\mu))
\ee 
where the first sum is over the data and the second is over a Monte
Carlo sample representive of the background. This method is not as
statistically optimal as using a proper likelihood -- however, it is still unbiased.

The weights $w^{(i)}_{\rm bkgd}$ account for the normalization difference
between the background in the data and in the Monte Carlo.  They are
computed in the manner indicated by Eqs.~(\ref{eq:wpeak}) and
(\ref{eq:wcombo}) in Sec. \ref{sec:back} below.  We call this
technique for handling the background the `direct unbinned background
subtraction' (DUBS) method.

The last term in Eq.~(\ref{pseudoll}) involves a sum
(Eq.~(\ref{apprxintegral})) over the Monte Carlo data sample. This
could be quite computationally intensive. However,
we are able to speed up this computation by transforming the sum over
events into a sum over `moments.'

Since our PDF can be written in the following form,

\be
\label{factorizedpdf}
F(\tilde x;\mu)=\sum_\alpha{A_\alpha(\mu)\times\Xi_\alpha(x)},
\ee
{\it i.e.}, as sum over a product of terms depending only on the fit
parameters and terms depending only on the kinematic variables, we
can define moments $M_\alpha$ by

\be
\label{moments}
M_\alpha=\frac{1}{N_{gen}}\sum_i {\frac{\Xi_\alpha(\tilde x_i)}{F(\tilde x_;\mu_{\rm mc})}}
\ee
where the sum is over reconstructed MC events, i.e., the same sum that defines $N(\mu)$ 
in Eq. (\ref{apprxintegral}).
This allows us to write $N(\mu)$ as a sum over moments:

\be
\label{momentsum}
N(\mu)=\sum_\alpha{A_\alpha(\mu)\times M_\alpha}.
\ee 
The moments can be computed once before fitting. In the fit, the time
consuming event sum is replaced with the sum over moments. In our case taking the
expansion of $h_{A_1}$ to order $(w-1)^2$ we have 18 moments to compute and sum. 

\subsection{Corrections and errors}

The approximations outlined above (the REC and DUBS methods) provide us with a
fast and easily implemented fitting procedure that uses the MC sample 
without any need to extract a detailed model of the
efficiency and resolution functions.  However, these advantages do not
come without a price: we must make corrections for the small residual bias 
and we must account for the uncertainty introduced by the REC and
DUBS procedures.

The REC method would be unbiased if the parameters $\mu_{\rm mc}$ used to
generate the Monte Carlo sample turned out to be the same as the true
parameters. This is, of course, in practice almost never the case, so we must
{\it ex post facto} introduce a small correction.  To do this we need
the derivatives of the fitted parameters with respect to the input
parameters. We obtain these by reweighting the Monte Carlo sample to
correspond to parameters that deviate by $\pm 0.1$ from the parameters used in
generation ($R_1=1.18$, $R_2=0.72$, $\rho^2=0.92$) and observing the
differences produced in the fitted parameters. This yields the full
$3\times 3$ matrix of the derivatives $d\mu_{\rm fitted}/d\mu_{\rm
true}$.

Given the derivative matrix we can relate the results of our fit to
the corrected results ($\mu_{\rm corrected}$) as follows

\be
\label{derivcorr}
(\mu_{\rm fitted}-\mu_{\rm mc})_=\frac{d\mu_{\rm fitted}}{d\mu_{\rm true}}\times (\mu_{\rm corrected}-\mu_{\rm mc}).
\ee
To solve for the corrected parameters we need only invert the
derivative matrix and multiply it against the LHS. The size of this
correction is $\sim 0.05$ for $R_1$, $\sim 0.07$ for $R_2$ and $\sim
-0.04$ for $\rho^2$.

Now since we use the theoretical PDF in our fit without explicit
smearing to account for resolution, we have also underestimated the
error. While the REC method corrects the values of the parameters for
resolution effects (to a first approximation), it does not account for
the loss of resolution due to our imperfect estimates of the four
kinematic variables (which arise largely from not being able to obtain
the exact $B$ rest frame, due to the missing neutrino, but also from
detector smearing).  However, the error matrix $E_{\rm fitted}$ 
for $R_1$, $R_2$ and $\rho^2$ can be
corrected using the derivative matrix, as follows

\be
\label{ecorrected}
E_{\rm fitted}\longrightarrow \lpar \frac{d\mu_{\rm fitted}}{d\mu_{\rm true}} \rpar ^{-1}E_{\rm fitted}\lpar \lpar\frac{d\mu_{\rm fitted}}{d\mu_{\rm true}}\rpar^{-1} \rpar^\dagger.
\ee
This increases the error estimates by $\sim 12\%$.

We call this the `bias map correction' (BMC) procedure, which refers
to both the modifications of the central values, and to
increase in the estimated statistical uncertainty.

In addition to the BMC error increase, there are contributions to the
error that are not accounted for in the fit. Two of these are Monte
Carlo statistical in nature: the error from the DUBS subtraction and
the error from the Monte Carlo integration used to evaluate $N(\mu)$
in the REC procedure. There is also a contribution to the statistical
error from the fluctuations in the background that is not accounted
for in the fit. The first two can be reduced by increasing the size of
the Monte Carlo sample, but the latter error is irreducible and must
be included regardless of the size of the Monte Carlo sample.

We have developed procedures for evaluating these three errors using
appropriate sums of the derivatives of the likelihood that can be
evaluated using the Monte Carlo sample. The formulas used are
collected in \Appx\ A. 

\label{sec:back}
\subsection{Background}

We rely on Monte Carlo to model the background. However, in order to
subtract the background it must be properly normalized. To do this we
adopt the background estimates from our $V_{cb}$
analysis~\cite{babarVcb}. The $V_{cb}$ analysis estimates the
background by fitting the $\cos\theta_{BY}$ and $\delta m$
distributions.  The $V_{cb}$ analysis fits the $\cos\theta_{BY}$
distributions for seven different background contributions (obtained
either from MC or data sidebands, see~\cite{babarVcb} for a
description of the seven categories). For our purposes we lump the
background into two categories -- peaking (candidates that contain a
properly reconstructed $D^*$ and hence peak in the $\delta m$
distribution) and combinatoric (candidates in which the $D^*$ is a
fake made from an incorrect combination of particles). Combinatorial
is one of the seven categories above; we obtain the peaking fraction
by summing over the other six.

The fractions of peaking and combinatorial background are determined
in ten $w$-bins.  Thus, we do not rely on the MC to model the
$w$-dependence of the background, but take it from data.

For each $w$-bin we compute weights to apply to each event from the MC
background sample before subtracting it.  For the peaking component
the weight for event $i$ is

\be
\label{eq:wpeak}
w^{(i)}_{\rm peaking}=f_{\rm peaking}(w)\times N_{\rm data}\times \frac{w_i}{\sum w_k}
\ee
where the $w_i$ are weights we apply to correct for difference in the
numbers of $B^0\bar B^0$, $B^+\B^-$ and continuum Monte Carlo events
available and $N_{\rm data}$ is the number of events in our signal
window. The sum is over peaking background events ($k$) that are in
the $w$-bin where the $w$ of the candidate falls.

This procedure guarantees that the number of peaking background events
subtracted will be equivalent to the number ($f_{\rm peaking}(w)\times N_{\rm data}$)
estimated to be inside the signal window.

A similar prescription applies for the combinatoric background with weights
defined by
\be
\label{eq:wcombo}
w^{(i)}_{\rm comb}=f_{\rm comb}(w)\times N_{\rm data}\times\frac{w_i}{\sum w_k}
\ee
where the events $k$ are now selected from the combinatoric category.

\section{Results}
\label{sec:Physics}

For the baseline result we perform the fit taking $R_1$ and $R_2$ to
be independent of $w$. All results are preliminary.

We find

\bea
\label{eq:r1r2rhosq}
\rone=1.328\pm 0.055\pm 0.025 \nonumber \\
\rtwo=0.920\pm 0.044\pm 0.020  \\
\rhosq=0.769\pm 0.039\pm 0.019 \nonumber
\eea
where the first error is statistical and the second Monte Carlo
statistical. The BMC (Eqns.(\ref{derivcorr}-\ref{ecorrected})) has been applied to both the central values and
the the \statl\ error (it shifts the former and expands the latter).  Systematic uncertainties are
discussed in Sec.
\ref{sec:Systematics} below.

The errors are highly correlated. 
The error matrix for the full statistical error for \rone, \rtwo, and \rhosq\
(including Monte
Carlo) is:

\bcenter
\begin{verbatim}
                             0.003664   -0.001696    0.001213 
                            -0.001696    0.002342   -0.001634 
                             0.001213   -0.001634    0.001850 
\end{verbatim}
\ecenter

The correlations are

\bea
\rho_{R_1-R_2}= -0.58  \nonumber \\
\rho_{R_1-\rho^2}= + 0.47 \\
\rho_{R_2-\rho^2}= -0.79  \nonumber.
\eea

As we do not yet have enough sensitivity
to fit for $w$-dependence of $R_1(w)$ and $R_2(w)$, we consider instead
the effect of the theoretically predicted dependence on the result.
Parameterizing this dependence as follows

\be
\label{r1w}
R_1(w)=R_1+\alpha_1(w-1)+\beta_1(w-1)^2
\ee
\be
\label{r2w}
R_2(w)=R_2+\alpha_2(w-1)+\beta_2(w-1)^2
\ee
and inserting these $w$-dependent forms into the PDF (with fixed $\alpha_i, \beta_i$ from 
the theoretical predictions) and fitting for the constant terms \rone\ and \rtwo\, we find the
results given in Table \ref{table:theorydep}.

\begin{table}[ht]
\begin{center}
\parbox{0.85\textwidth}{\caption{\label{table:theorydep}
Dependence of form factor parameters on 
theoretical assumptions about slope ($\alpha$) and curvature ($\beta$) of $R_1$ and
$R_2$ $w$-dependence. See Eqns. (\ref{r1w}) and (\ref{r2w}).
}}
\begin{tabular}{ |c|l l l l l l l|} 
\hline\hline
 Reference\hfill & \ $\alpha_1 $ & $\beta_1$ & $\alpha_2$ & $\beta_2$
 & $R_1$ & $R_2$ & $\rho^2$ \\ \hline\hline
 Neubert\cite{NeubertPhysReport} & -0.22 & 0.09 & 0.15 & -0.04                      & 1.37 & 0.88 & 0.75 \\
\hline
Caprini-Lellouch-Neubert\cite{Caprini}       & -0.12   &     0.05     &      0.11   & -0.06 & 1.35 & 0.89  & 0.76 \\
\hline
Ligeti-Grinstein\cite{LigetiGrinstein}    & -0.10  &     0.0      & 0.09         & 0.0 & 1.35 & 0.89 & 0.75 \\
\hline
Baseline                                  & 0.0   &     0.0      & 0.0          & 0.0 & 1.33 & 0.92 & 0.77 \\
\hline\hline
\end{tabular}
\end{center}
\end{table}

Taking into account these theoretical variations in $R_1(w)$ and $R_2(w)$ yields 
slightly larger values for $R_1$ and slightly
smaller values for $R_2$. The prediction by Neubert produces more deviation than the more recent calculations.

\section{Goodness-of-fit}
\label{sec:gof}
As these results are obtained from a maximum likelihood fit, we need
some method of assessing whether the results of the fit actually
reproduce the distribution of the kinematic variables in the data.
Since we do not have an explicit form for the acceptance-corrected PDF
($\tilde F$) of the four reconstructed variables $w$, $\ctl$, $\ctv$
and $\chi$, we reweight the MC sample to construct the distributions
expected from our measured parameters. That is, the contribution of
the signal to a bin is

\be
\label{eq:wtsum}
n_{\rm signal}=\sum{w^{(i)}_{\rm signal}}
\ee
where 
\be
w^{(i)}_{\rm signal}=\left(1-f_{\rm peaking}(w)-f_{\rm signal}(w)\right)\times N_{\rm data}\times\frac{w_i}{\sum w_i}
\ee
and in this case $w_i=\frac{F(x;\mu)}{F(x;\mu_{mc})}$ is the weight
needed to modify the distributions from those generated with
$\mu_{mc}$ ($R_1=1.18$, $R_2=0.72$, and $\rho^2=0.92$), to those we
obtain from this analysis.

For the background we use the same weighting procedure (see
Eqs.~(\ref{eq:wpeak}) and (\ref{eq:wcombo})) used in the fit. Using
these weighting procedures the normalizations of the data and
reweighted distributions match by construction.

We consider two types of goodness of fit: (a) five one-dimensional
distributions (the projections $w$, $\ctl$, $\ctv$, and $\chi$ plus
the distribution of CM lepton momentum $p^*_\ell$) and (b) a binned
$\chi^2$ based on $6\times 6\times 6\times 6$ (a total 1296) bins. For
a comparison we also give the unweighted results which corresponds
closely to those using the CLEO parameters~\cite{CLEO}.

In Figs.~\ref{fig:cleoplots}-\ref{fig:pstarlplot} a comparison of the
projections in the kinematic variables and $p^*_\ell$ between the
Monte Carlo and the data can be seen.  The plots in Fig.~\ref{fig:cleoplots}
show the result for the default parameters, while those in
Fig.~\ref{fig:ourplots} give that obtained by reweighting by our
parameters, and Fig.~\ref{fig:pstarlplot} shows both for the
$p^*_\ell$ spectrum. The fit is to a constant 
from which we read off the $\chisq$.  

Close inspection will reveal that the ratio plot projections in
Fig.~\ref{fig:ourplots} show better agreement to the line at unity
than those in Fig.~\ref{fig:cleoplots}, but the improvement is more
clearly seen numerically in Table \ref{tbl:chisqs}, which gives the
$\chisq$ for fitting each plot to a constant. The constant is always
consistent with unity as it must be.  In every case the agreement
between MC and data improves when we use our result -- sometimes
substantially.

\begin{table}[ht]
\begin{center}
\parbox{0.85\textwidth}{\caption{\label{tbl:chisqs} $\chisq$ and $\chisq$-probability 
for kinematic variable projections and lepton momentum; the number of bins  in these 
histograms is 10. The number of degrees-of-freedom is 9, since we have
forced the normalization.}}
\begin{tabular}{|c|c|c|}
\hline\hline
variable & $\chi^2$ (prob.) default & $\chi^2$ (prob.) BaBar   \\
\hline\hline
$w$ & 14.08 (0.120) & 13.69 (0.134)  \\
\hline
$\ctl$ & 17.12 (0.047) & 7.626 (0.572) \\
\hline
$\ctv$ & 40.93 (0.000) & 17.81 (0.0374)  \\
\hline
$\chi$ & 8.133 (0.521) & 7.082 (0.623)  \\
\hline
$p^*_\ell$ & 17.88 (0.037) & 7.316 (0.604)  \\
\hline\hline
\end{tabular}
\end{center}
\end{table}

The value of the four-dimensional binned $\chisq$ goes from $1274.04$
with default MC parameters to $1232.23$ with our \param s -- an
improvement of \approxx $42$ units of $\chi^2$. The $\chi^2$ per bin
is slightly smaller than unity. It is not really proper to interpret
this as a probability as there are about 200 empty bins which
contribute nothing to $\chisq$.  Nonetheless it is clear that the
reweighted MC follows the distribution of the data quite well and that it models this 
decay (the single largest $B$ branching fraction) much better than the default MC.

\clearpage
\begin{figure}[hb]
\begin{center}
\begin{tabular}{ c c }
{\parbox{8cm}
  {\resizebox{!}{8cm}{\includegraphics{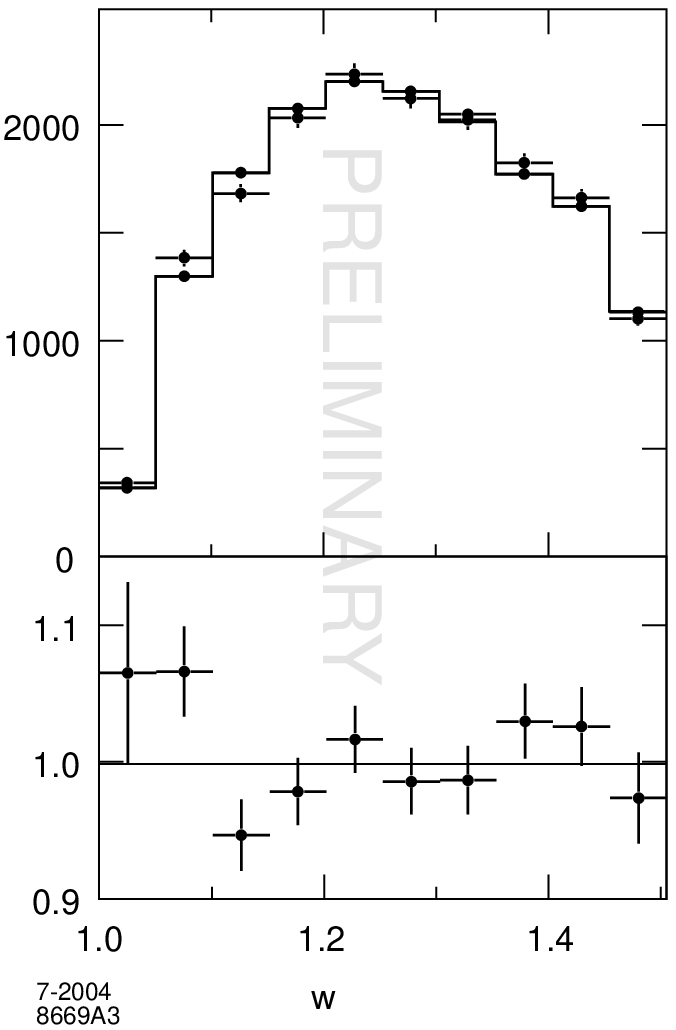}} 
}}
&
{\parbox{8cm}
  {\resizebox{!}{8cm}{\includegraphics{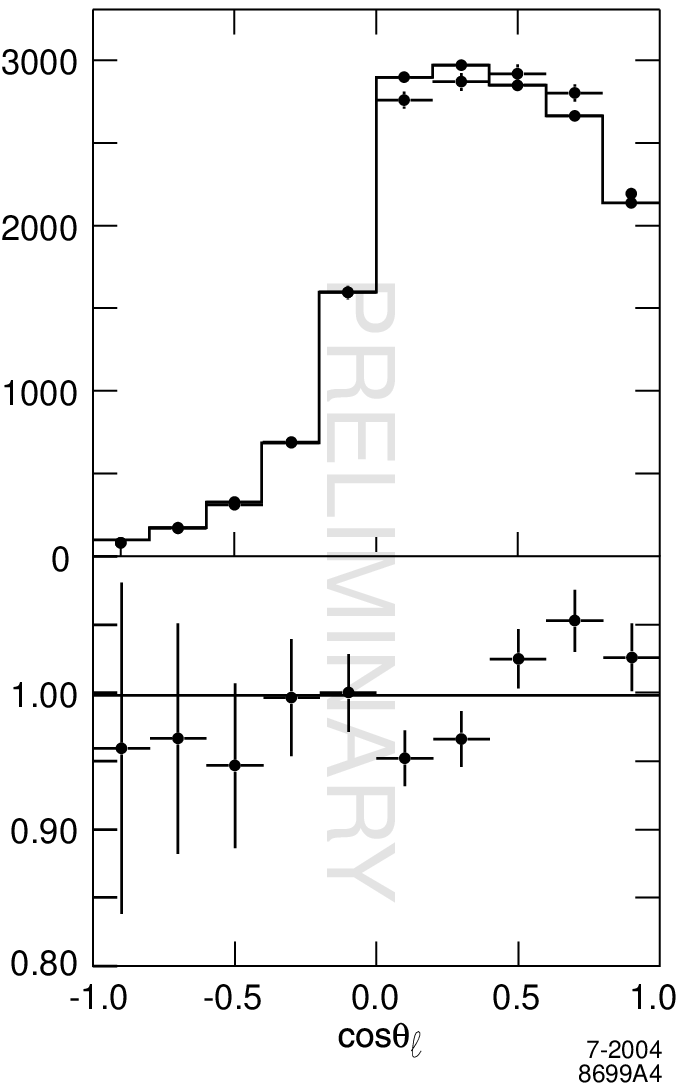}} 
}} \\
{\parbox{8cm}
  {\resizebox{!}{8cm}{\includegraphics{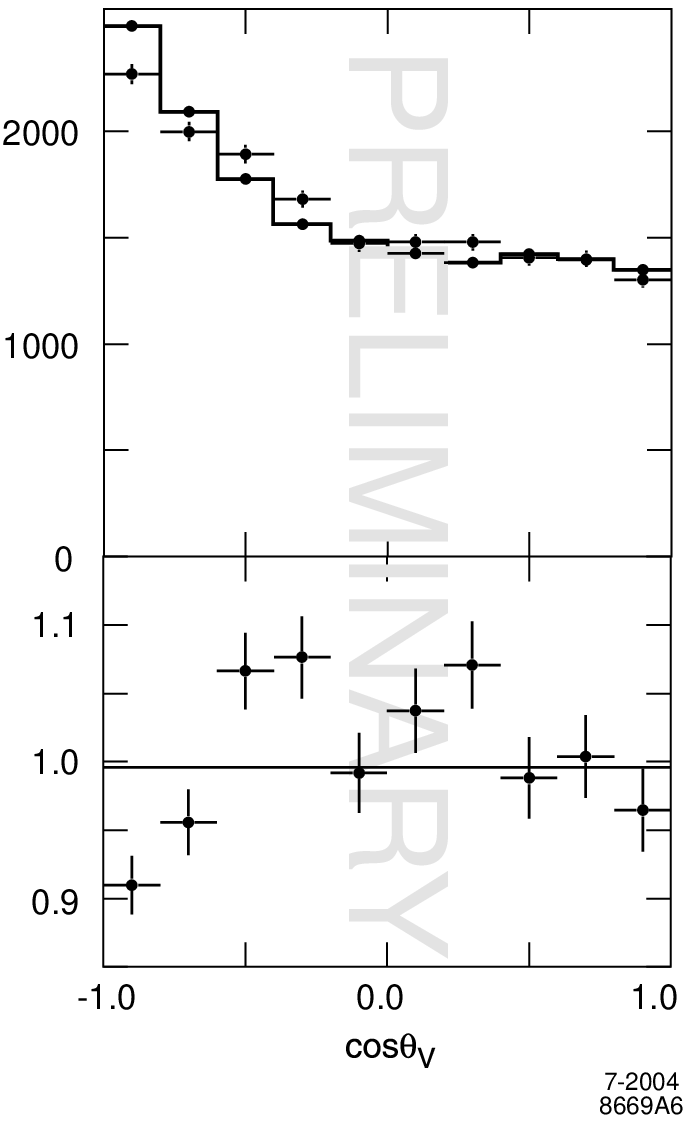}} 
}} 
&
{\parbox{8cm}
  {\resizebox{!}{8cm}{\includegraphics{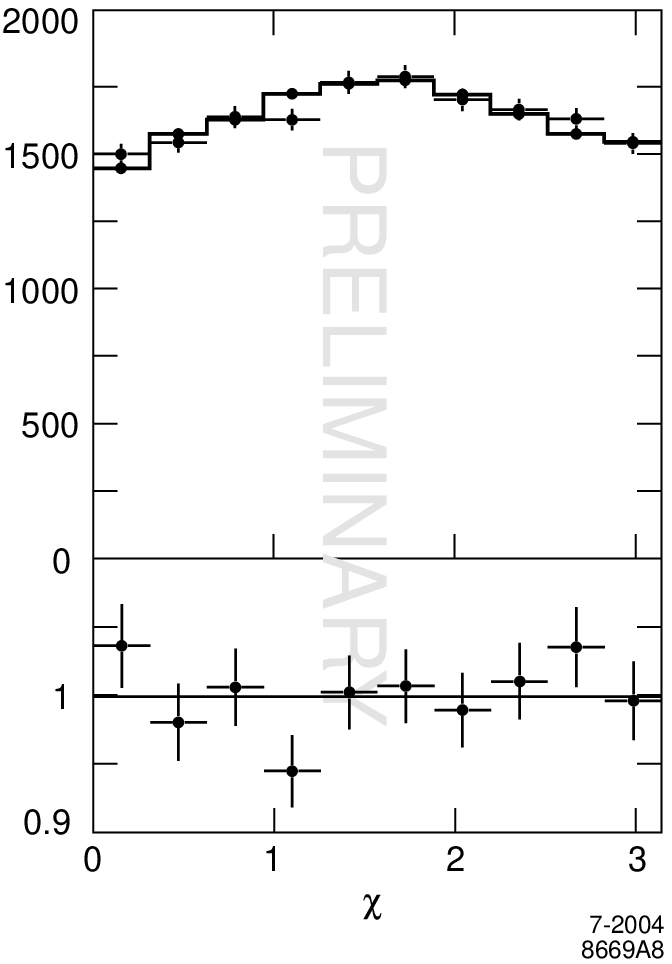}} 
}} \\
\end{tabular}
\end{center}
\caption{\label{fig:cleoplots} Data (points) overlayed on Monte Carlo (histograms) 
for all four \kv\ distributions for default values of the parameters 
The plots below the overlays are the ratios of the data to MC distributions.  }
\end{figure}

\clearpage
\begin{figure}[hb]
\begin{center}
\begin{tabular}{ c c }
{\parbox{8cm}
  {\resizebox{!}{8cm}{\includegraphics{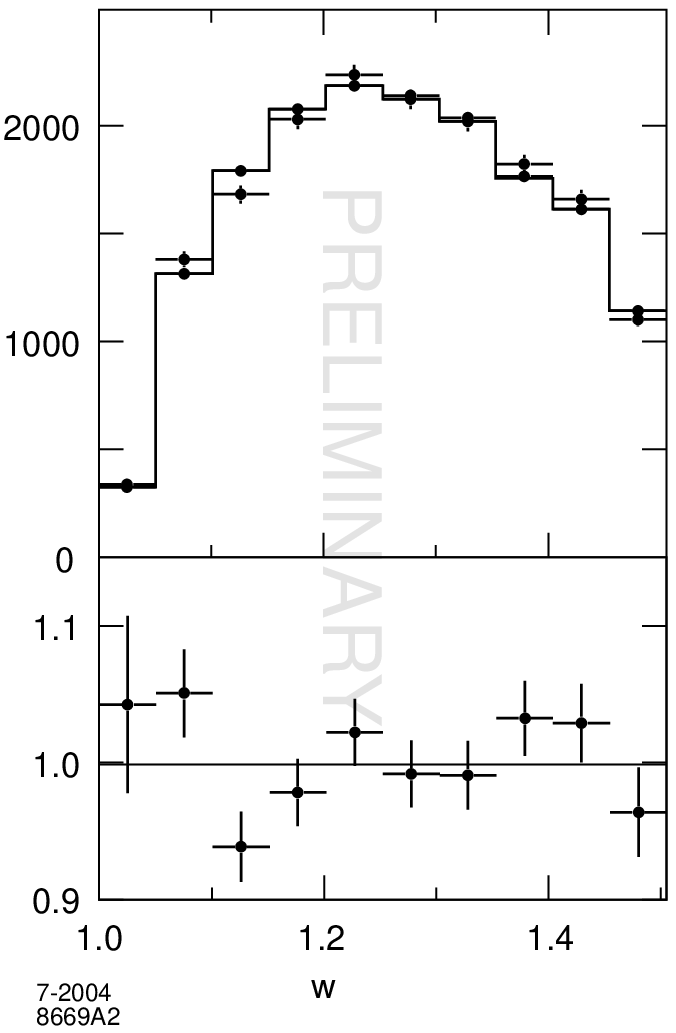}} 
}}
&
{\parbox{8cm}
  {\resizebox{!}{8cm}{\includegraphics{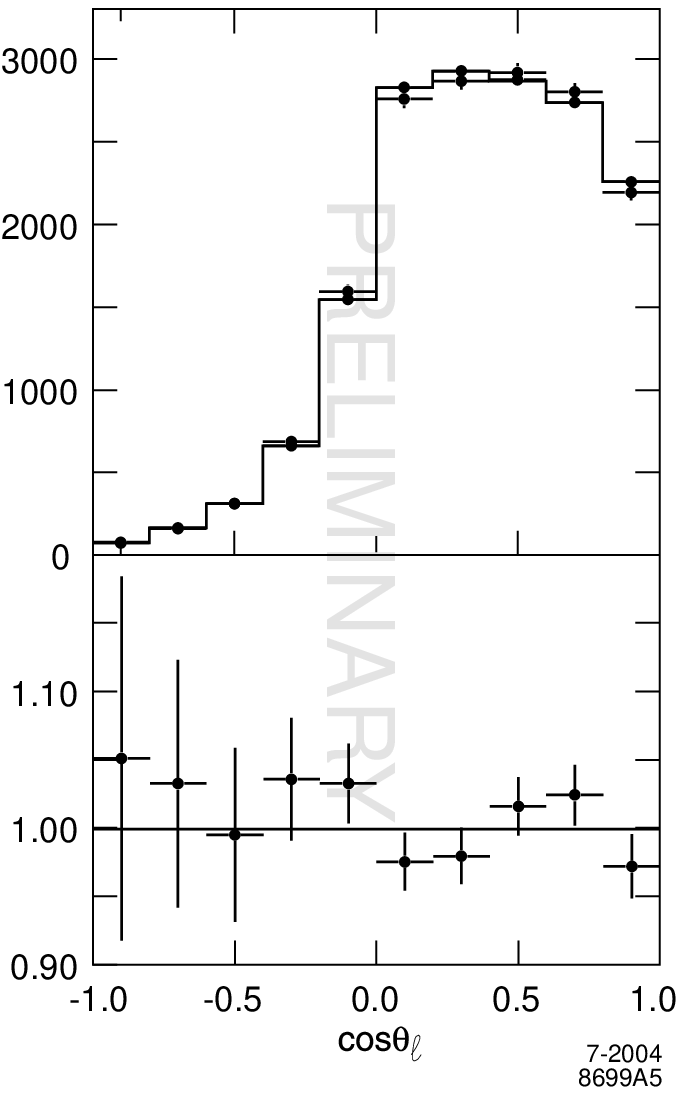}} 
}} \\
{\parbox{8cm}
  {\resizebox{!}{8cm}{\includegraphics{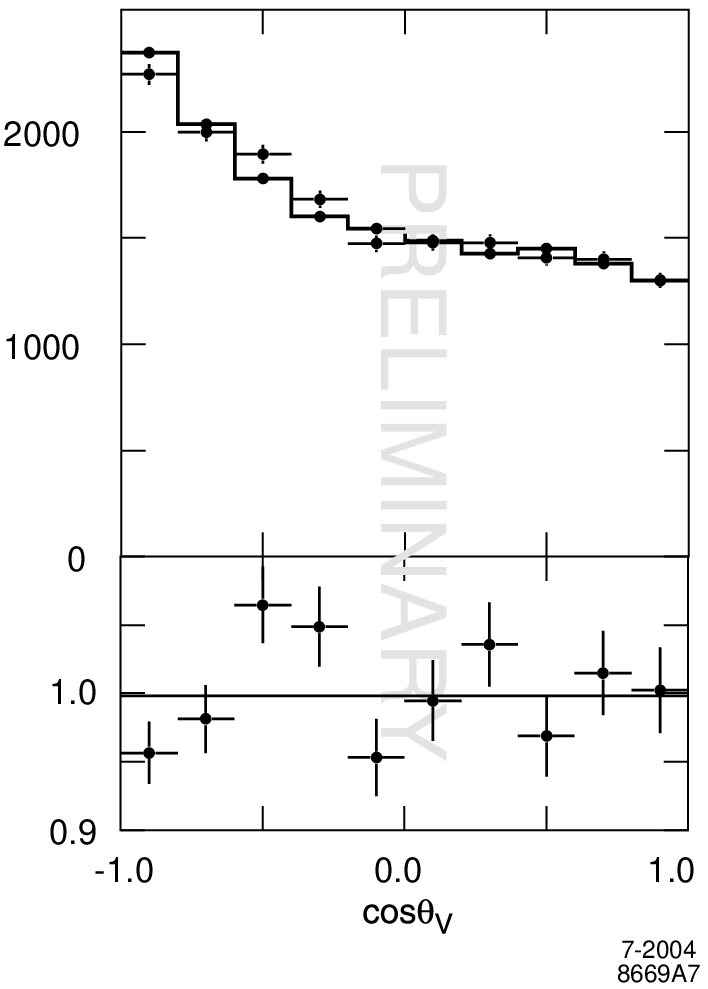}} 
}} 
&
{\parbox{8cm}
  {\resizebox{!}{8cm}{\includegraphics{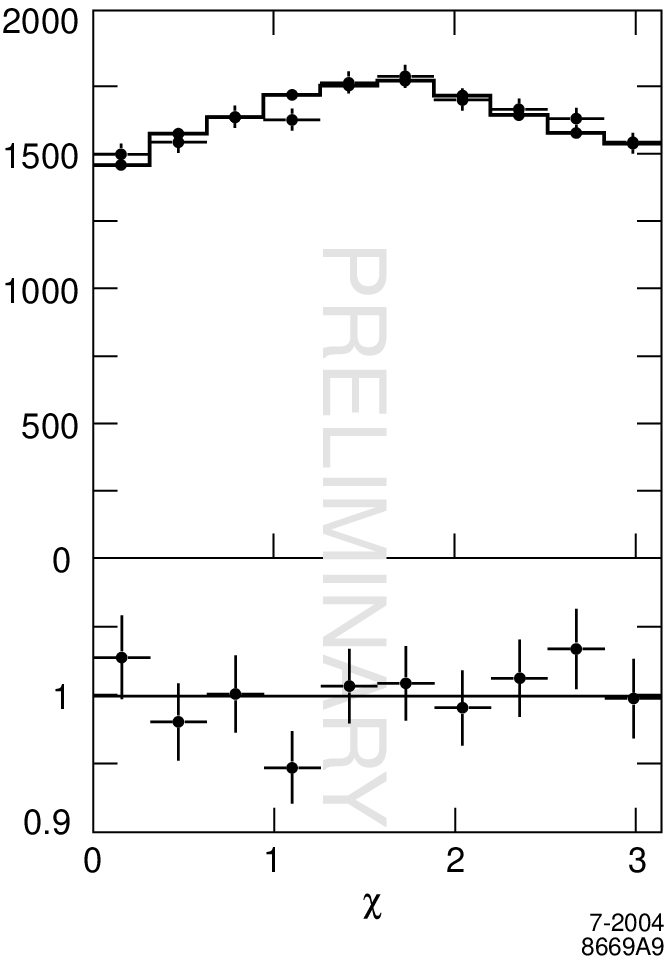}} 
}} \\
\end{tabular}
\end{center}
\caption{\label{fig:ourplots} Data (points) overlayed on Monte Carlo (histogram) 
for all four \kv\ distributions use our measurement of the parameters 
The plots  below the overlays are the ratio of the data to the MC distributions.  }
\end{figure}

\clearpage\begin{figure}[hb]
\begin{center}
\begin{tabular}{ c }
{\parbox{10cm}
  {\resizebox{!}{10cm}{\includegraphics{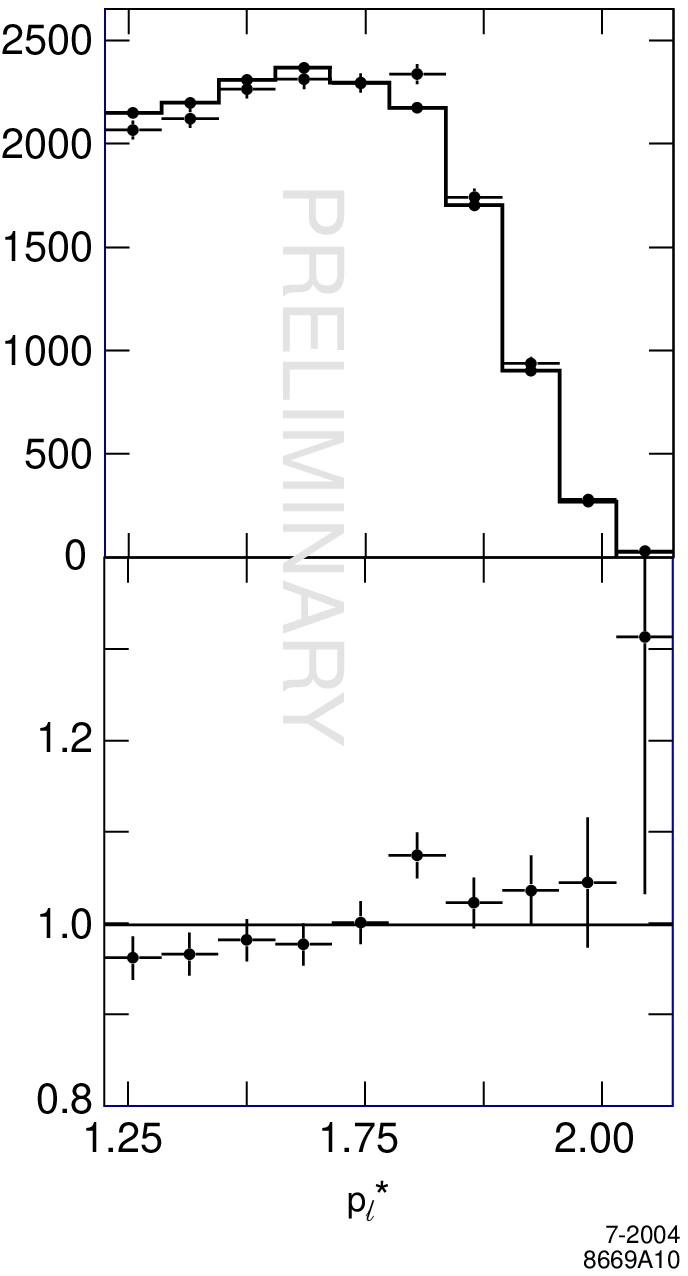}}
}} \\
{\parbox{10cm}
  {\resizebox{!}{10cm}{\includegraphics{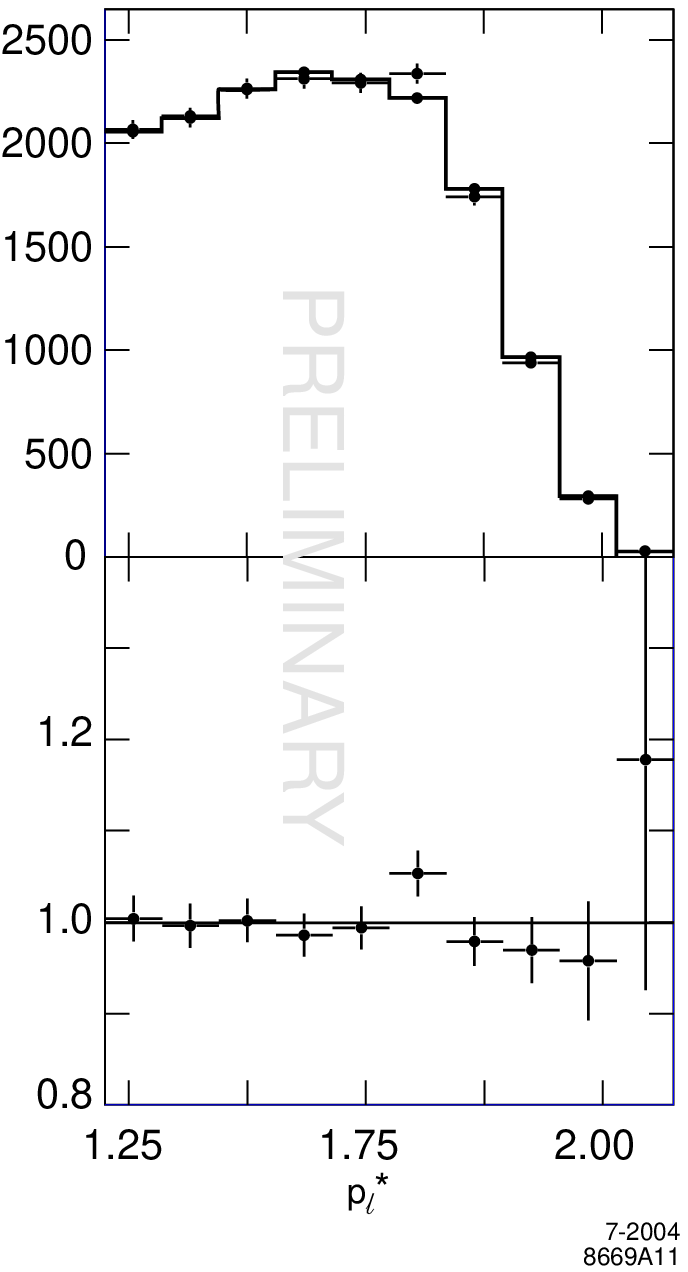}}
}}
\end{tabular}
\end{center}
\caption{\label{fig:pstarlplot} Data/Monte Carlo comparison for $p^*_\ell$-distribution 
(data are points, histograms are MC).  MC used for the top plot 
uses default parameters and MC used for the bottom is reweighted to our parameters.}
\label{fig:plep-plot}
\end{figure}

\section{Systematic Studies}
\label{sec:Systematics}
The systematic uncertainties on the three principal parameters are
summarized in Table~\ref{table:systerrs}.  The dominant systematic
errors arise from the MC simulation: that is, from how well we
understand and simulate the detector performance in terms of
resolution and efficiencies, in particular the efficiency for the
reconstruction of low-momentum
 charged pions from \Dstarplus\ decays.
Further, how well we model the signal and background event generation,
{\it e.g.} how close the branching fractions in the event generator
are to those of the real world, affects the distribution of background
we subtract. 
The $w$-dependence of the
background, however, is not taken from the MC but is measured in
the data and the uncertainty in this measurement also contributes
to the systematic errors.

\subsection{Detector Simulation}
Extensive studies of the simulation of the detector response,
including careful examination of track reconstruction efficiencies
and particle identification, have been performed using selected data
control samples. 
Adjustments for
known simulation deficiencies are used in 
 investigating and evaluating
 the
systematic errors.  
 Form factor measurements are insensitive to
overall normalization errors.  
 Thus differences in the efficiencies
that are independent of the fit variables do 
 not affect the results,
but variations of differences as a function of these variables are of
concern.  

To assess the uncertainties due to differences in shape
rather than normalization 
 we vary the dependence on the efficiency
corrections, reprocess the MC samples, redo 
 the fit to the data, and
take the difference to the results obtained with the nominal 
 MC
simulation as an estimate of the systematic error on the parameters.
This procedure is repeated for every source of systematic
uncertainty. 
 The individual uncertainties are added in quadrature to
obtain the total systematic error.

\subsubsection{Charged Particle Tracking}
\label{sec:trking} 

The difference between the tracking efficiency for electrons, and charged kaons and pions from $D$ decays decreases roughly linearly as a function of momentum.  We vary this linear dependence on the particle momentum in the MC simulation and observe no significant change in the results. Consequently, we assign no error from this source.

\subsubsection{ Slow Pion Reconstruction }

The efficiency for reconstructing the low-momentum charged pion ($\pi_s$) from $D^{*+}$ is a 
major source of systematic errors. Because of the small energy release in \Dstar\ decays, 
this pion is emitted in the same direction as the parent \Dstar\ and its momentum is less 
than 400 \mevc\ in the laboratory frame.  
Since $w=E_{D^*}/m_{D^*}$ in the B rest frame, the $\pi_s$ momentum is correlated with $w$ 
and thus its momentum dependent efficiency impacts the measurement of $\rho^2$. 

The uncertainty due to the low-momentum tracking efficiency is evaluated differently 
from other tracking errors because such low-momentum tracks do not traverse the whole drift chamber.  
Their detection and measurement depends mostly on the silicon vertex tracker. 
To study this efficiency as a function of $p_{\pi_s}$ we 
use a large set of $\dsp\ra\Dz \pi_s^+$ decays selected from hadronic events and 
measure the 
distributions of the helicity angle of $\pi_s^+$ ($\theta_V)$ in the $D^{*+}$ rest 
frame as a function of the $D^{*+}$ momentum.
Fig.~\ref{fig:dsthel} in Sec.~\ref{sec:mc} compares the $\cos\theta_V$ obtained  
for data and MC simulation. The small size of the linear contributions to the
correction factors (Eqn. \ref{eq:fcorr}) is encouraging, but there is enough room
in the error that we still need to include the $\pi_s$ efficiency in the systematic error.

We parameterize the effective $\pi_s$ efficiency as function of its momentum using the form
\be
\label{eq:epis}
\varepsilon (p_{\pi_s})=\varepsilon_{\rm max}\left(1-\frac{1}{1-\beta(p_{\pi_s}-p_0)}\right)
\ee
with $p_0$ being the threshold momentum and $\beta$ controling the rapidity 
with which the efficiency rises above threshold.
For $p_{\pi_s}<p_0$ we set $\varepsilon$ to zero.
We fit the data and MC helicity angle distributions for $\beta$, $p_0$
and the coefficient of
 the physically allowed $\cos^2\theta_V$ term
to obtain efficiency functions for the data and the MC. 
The helicity method only determines the relative momentum dependence of the efficiency, so the
normalization $\varepsilon_{max}$ must be determined separately. Since, 
normalization does not
matter for this analysis we simply set it to unity.

To assess the systematic uncertainty due
to the $\pi_s$ efficiency we weight the MC simulation by the ratio
of data to MC functions and 
 assign the observed shifts in the fitted
values for $R_1$, $R_2$ and $\rhosq$ as systematic errors.  
 Not
unexpectedly, $\rhosq$ is most sensitive to this efficiency since 
 it
describes the shape of the $w$-distribution.

\subsubsection{Charged Particle Identification (PID)}

Using data and MC simulated control samples we have tabulated  the difference in 
particle identification efficiency for data and MC. These tables provide the
correction factors in bins of momentum and angle. Since this analysis is most
sensitive  to efficiency variations with momentum
we average 
the tables over the angles to obtain corrections as a function of momentum. 
For electrons the correction factors  vary 
from 0.991 to 1.008 over the momentum range from $1.2 \gevc$ to $2.5 \gevc$.
We assess the impact of the uncertainty in these 
corrections by approximating their momentum dependence 
by linear functions and vary the sign of the small slope of these functions.  The observed 
deviations from the default fit are $\Delta R_1=0.0064$, $\Delta R_2=0.0052$
and $\Delta \rhosq=-0.0016$ for the positive slope and -0.0032, -0.0031, +0.0009 for
the negative slope. We take half of the difference as the systematic error from this
source.  Since the momentum dependence is not a monotonic function, this procedure overestimates the uncertainty.

For kaon identification we employ the same procedure. The observed variations are significantly smaller.

The probability of misidentifying charged hadrons, $\pi^{\pm}$, $K^{\pm}$, $p^{\pm}$, 
as electrons is very small, less than $0.2\%$ in the momentum range $1.2-2.5 \gevc$.  
Since a variation of the peaking background by 9\% results in a very modest change in 
the fit results, and since the fraction of this background originating from hadrons 
misidentified as electrons is very small, we
conclude that the uncertainty in the hadron misidentification rate should be negligible.

The misidentification rate of pions as kaons ranges from a few tenths of a 
percent  to almost $5\%$. However, pion misidentification is well simulated by
the MC and thus should have little impact on the fit results. Furthermore, the main 
consequence of pion misidentification is to increase the combinatorial background. Since we estimate 
the combinatorial background from a fit to the measured $\delta m$ distribution, we are not 
dependent on the MC to assess the size of this background. We conclude that the 
uncertainty in the pion misidentification rate has little impact on the fit results.

\subsection{Event Simulation}

\subsubsection{Final State Radiation (FSR) }
\label{sec:fsr}

Final state radiation, primarily from electrons, lowers the momenta and 
to a lesser degree changes the angles of detected particles. 
Though a physics effect, FSR acts much like a resolution -- it smears the kinematic variables. 
We simulate the emission spectrum of radiative photons using PHOTOS~\cite{Was},
so the REC method  corrects for it, to the extent that PHOTOS models it correctly.

To test the sensitivity to FSR  we evaluate the shifts 
in the fitted values of $R_1$, $R_2$ 
and $\rhosq$  between fits done with and without FSR corrections.
We assume an uncertainty of 30\% in the simulated photon emission and 
thus take $\sim 1/3$ of the observed shifts of $0.0129$, $0.067$ and $0.0039$ as 
an estimate of the systematic uncertainty.  

\subsubsection{Background Simulation}
\label{sec:backall}

We divide the background sources into two categories: 
(1) peaking background for which the $D^*$ decay has been correctly 
reconstructed
and which contributes to the peak in the $\delta m=m_{K\pipi}-m_{K\pi}$ 
distribution and (2)
combinatoric background for which the $D^*$ has not been properly reconstructed and 
thus does not contribute to the peak in the $\delta m$ distribution.

\hfill
\par\noindent{\bf Background mixture  }

Our modeling of the  peaking and combinatorial background depends on the not
very well known branching fractions for the mixture of semi-leptonic $B$ decay modes that make up the 
background.
To estimate the uncertainty associated with these branching fractions, we vary their 
values and observe shifts in the fit parameters compared to the nominal values. In this 
process we keep the total background fractions as determined by the $\cos\theta_{BY}$ fit 
(see Sec. \ref{sec:back}) unchanged.
We vary most modes by $60\% $. For the measured mode $B\rar D_1l\nu$ we vary only by 
the $\sim 30\%$ measurement error\cite{pdg2002}.  
In the case of $D^*l\nu$ there are contributions from badly reconstructed signal events.  
We assume a $20\%$ uncertainty in the signal branching fraction to account for the 
large differences in the currently available measurements.

\begin{table}[hb]
\bcenter
\parbox{0.85\textwidth}{\caption{ \label{tbl:bsemilep}  Systematic errors due to  the uncertainties in the branching fractions of semileptonic $B$ decay modes. 
 The  branching fractions listed are the default values used in 
the MC simulation.
The isospin related modes of neutral and 
charged $B$ mesons are varied together by the factors indicated. }}
\begin{tabular}{lcclll}
\hline \hline
Decay Mode  &    MC branching fraction (\%)  &  Variation (\%) & $\sigma_{R_1}$& $\sigma_{R_2}$ &  
$\sigma_{\rho^2}$ \\
\hline
$B \rar D^* \ell \nu $    &  5.6  & 20 & 0.00052& 0.00044 & 0.00027  \\ 
$B \rar D \ell \nu$       &  2.10 & 20 & 0.0013 & 0.00037 & 0.0002   \\ 
$B \rar D_0 \ell \nu$     &  0.20 & 60 & 0.00020& 0.00026 & 0.00010  \\ 
$B \rar D_1 \ell \nu$     &  0.56 & 30 & 0.0087 & 0.0.0024& 0.0016   \\ 
$B \rar D'_1 \ell \nu$    &  0.37 & 60 & 0.012  & 0.0062  & 0.0044   \\ 
$B \rar D_2 \ell \nu$     &  0.37 & 60 & 0.00095& 0.0025  & 0.0017   \\ 
$B \rar D^*\pi \ell \nu$  &  0.30 & 60 & 0.0036 & 0.00087 & 0.00071  \\ 
$B \rar D\pi  \ell \nu$   &  0.9  & 60 & 0.0022 & 0.00092 & 0.00061  \\ 
\hline\hline
$e^+e^- \ra c\bar c$      & NA    & 20 & 0.0011 & 0.00034 & 0.00040  \\ 
\hline
Total                     &       &    & 0.016  &  0.0073 & 0.0051 \\
\hline\hline
\end{tabular}
\ecenter
\end{table}

In Table \ref{tbl:bsemilep} we list these branching ratio 
variations as well as the effect of varying the  contribution from $\eplus\eminus \ra c\bar c$ events. 
We take half of the observed variation in the fit parameters as an estimate for the systematic error. 

The total error for the three fit parameters is obtained 
by adding the errors due to each contribution in quadrature.
$R_1$ is the most sensitive to the mixture of decay modes of the background
subtraction. 

\hfill
\par\noindent{\bf Dependence of the Background on  $w$}

The $w$-dependence of the background estimate is taken from $\cos\theta_{BY}$ fits
performed for each bin in $w$~\cite{babarVcb}.
We fit the $w$-dependence of the peaking and the combinatorial backgrounds 
to a second order polynomial centered 
at the middle bin. We use these
polynomials, $f_{\rm peaking}(w)$ and $f_{\rm comb}(w)$, to compute the weights (see Eqs. (\ref{eq:wpeak}) and
(\ref{eq:wcombo})) that normalize our background subtraction.

To estimate
the error from the $w$-dependence we vary the slopes of  
$f_{\rm peaking}(w)$ and $f_{\rm comb}(w)$ polynomials 
by $\pm 1\sigma$ and refit for $R_1$, $R_2$ and $\rho^2$ with the altered
background normalizations.
We take half of the total variation from each background type as its
contribution to the systematic error. 

\hfill
\par\noindent{\bf Total peaking and \comb\ \bkgd\ fractions}

Another source of  error is the uncertainty in the \normzn\ of the
peaking and \comb\ background fractions.  To estimate this error we vary the fraction  $f_{\rm peaking}$ by $\pm 12\%$
and $f_{\rm comb}$  by $\pm 15\%$.
We have increased these uncertainties beyond the statistical uncertainites
established by our \vcb\ analysis \cite{babarVcb} to account for slight differences
between the way the analyses define backgrounds.
For  $f_{\rm peaking}$
this yields  errors of 0.0186, 0.0075, 0.0014 for the three parameters, 
indicating that this is  a significant error source for \rone.
For $f_{\rm comb}$ we find error estimates
of 0.0063, 0.0034, 0.0080.

\hfill
\par\noindent{\bf MC/data side-band comparison}

 The distributions of the kinematic variables for MC and data agree
well
 in the $\delta m$ sideband region used to estimate the
combinatoric
 background. But since the \comb\ \bkgd\ comprises about
a third of the
 total \bkgd\ under the peak, the small differences in
the shapes of
 the distributions could introduce an error in the
background subtraction process (when the MC is used to
 subtract the
residual \comb\ background from the data, see
Sec.~\ref{sec:fitting}). To estimate the impact of this effect, we
first take the ratios of data to MC in
 the \sbr\ and then fit the
distributions with polynomials with the
 result shown in
Fig.~\ref{fig:hsbDoMCratio}.

We then use these functions one at a time to multiply the
combinatoric
 background from MC before it is subtracted from the data
to prepare
 the sample for fitting.
We then carry through the fits and then take the differences in the \FF s we obtain from
 these with the
\FF s we obtain from the fits with the unaltered background. 
This procedure yields the results shown in
Table~\ref{rwtcombbkgd}.
 
The differences we find are small. The largest is from using
the function for \recow, from which we find $\Delta \rho^2$
$\sim 0.006$. Since in the end we take the $w$-dependence of the background
from the data and the deviations due to weighting the angular distributions 
are very small, we add nothing to
the systematic uncertainty from this check.

\begin{figure}[ht]
\bcenter
{\parbox{10.5cm}
{\rotatebox{0}{\resizebox{!}{10cm}{\includegraphics{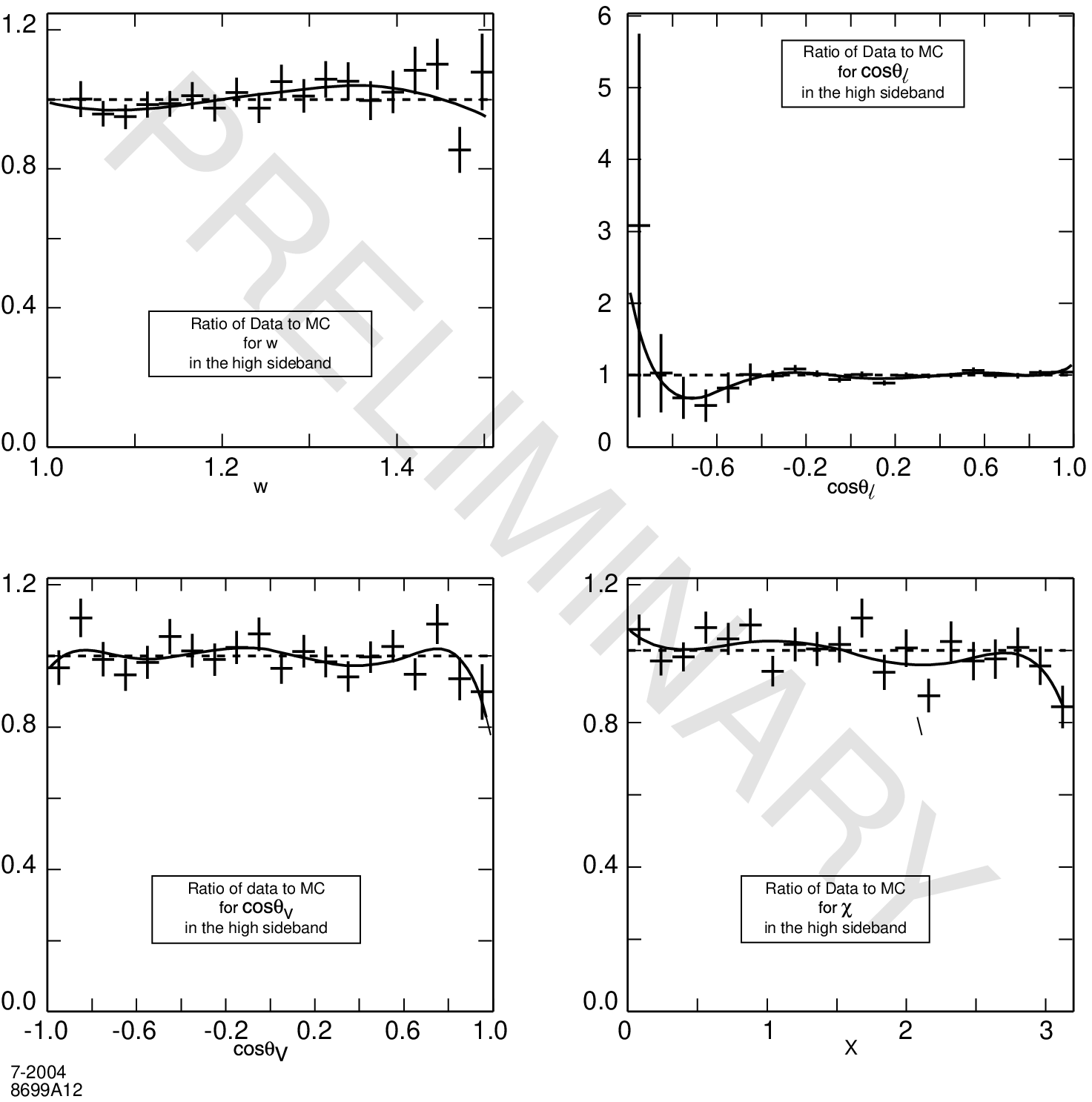}}
}
}
} 
\caption{Ratio of data to MC for the four \kv s in the high \sbr\  data, and the best polynomial fits.}
\label{fig:hsbDoMCratio}
\ecenter
\end{figure}


\vsp
\begin{table}[ht]
\bcenter
\parbox{0.85\textwidth}{\caption{\label{rwtcombbkgd}  Changes
in the fitted parameters for reweighting of the MC combinatoric
background distributions in the four kinematic variables, as shown in
Fig.~\ref{fig:hsbDoMCratio}.
}}
\begin{tabular}{lccc}
\hline \hline
Reweighted distributions & $R_1$ & $R_2$ & $\rho^2$ \\
\hline
w distribution     & -0.002 & 0.0 & 0.006 \\
\ctl\ distribution & 0.001 & -0.002 & -0.001 \\
\ctv\ distribution & 0.002 & -0.003 & 0.001 \\
\angchi\ distribution & 0.004 & -0.002 & 0.001 \\
\hline\hline
\end{tabular}
\ecenter
\end{table}

\subsection{ Summary of Systematic Errors  }

The systematic errors are summarized in Table
\ref{table:systerrs}. While the total error remains statistics-dominated, we are approaching the systematic 
limit. 
 The addition of
other $D^0$ decay modes with higher multiplicity and higher
backgrounds is likely  to increase our sensitivity to the various
background sources. Enhancing our understanding of these backgrounds will be
critical to improving the errors. 
 Measurements of the higher mass
semileptonic decay modes are sorely needed.

\begin{table}[ht!]
\begin{center}
\parbox{0.85\textwidth}{\caption{\label{table:systerrs}  Summary of the estimated \Syst\ errors.   Negligibly small 
contributions have been omitted here, they are discussed in the text: 
\Dzero\ tracking, FSR, and $\delta m$  MC/data differences. }
}
\begin{tabular}{ lllll }
\hline\hline
\em{Error source}        &  $\sigma_{\rone}$ & $\sigma_{\rtwo}$ &  $\sigma_{\rhosq}$  \\
\hline
Lepton-hadron track efficiency  & 0.005 & 0.004 & 0.002   \\  
Slow pion track efficiency      & 0.003 & 0.0002 & 0.011  \\
PID misID (lepton, kaon)        & 0.005 & 0.004  & 0.002  \\
Peaking background normalization        & 0.019 & 0.008  & 0.001  \\
Combinatorial background normalization  & 0.006 & 0.003 & 0.008  \\  
Background composition (branching fractions)     & 0.016 & 0.0079 & 0.0051  \\
$w$-dependence of background    & 0.001 & 0.001  & 0.028  \\  
Final state radiation           & 0.0043& 0.0023 & 0.0013 \\
\hline 
Total  \Syst\                   &  0.027 & 0.014 &  0.033   \\
\hline\hline
\end{tabular}
\end{center}
\end{table}

\section{Conclusions}
\label{sec:Summary and disscussion}
We have measured the form factors $A_1$, $V$ and $A_2$ in terms of the
HQET-inspired parameters $R_1$, $R_2$ and $\rho^2$. Note that
all results are preliminary.

The baseline result
including systematic errors is

\be
\label{eq:resultwithsys}
R_1=1.328\pm 0.055\pm 0.025\pm 0.025, 
\ee
\bes
R_2=0.920\pm 0.044\pm 0.020\pm 0.013,
\ees
\bes
\rho^2=0.769\pm 0.039\pm 0.019\pm 0.032
\ees
where the first error is \statl, the second MC \statl, 
and the third \syst. This
baseline result is obtained neglecting any possible 
$w$-dependence of $R_1(w)$ and $R_2(w)$. 
It agrees well, but not perfectly, with theory at $w=1$ (see eq.~(\ref{eq:neubr1r2}) or 
(\ref{eq:caprini}) ). We achieve a considerable improvement over the CLEO result of
$R_1=1.18\pm 0.30\pm 0.12$, $R_2=0.71\pm 0.22\pm0.07$ 
and $\rho^2 = 0.91\pm 0.15\pm 0.06$ \cite{CLEO}.

We do not yet have the sensitivity to independently establish the
$w$-dependence of $R_1(w)$ and $R_2(w)$, but we can use theoretical
estimates to make comparisons.  If we compare the predictions of 
Caprini, Lellouch and Neubert~\cite{Caprini} for $R_1$ and $R_2$ to
the result we obtain using their $w$-dependence (see Table
\ref{table:theorydep}), we find $\Delta R_1=1.36-1.27=0.09$ ($1.6\sigma$
statistical) and $\Delta R_2=0.87-0.80=0.07$ ($1.4\sigma$
statistical). Given the undoubted presence of theoretical errors, this
is reasonable agreement.

This measurement allows us to make a considerable improvement in the size of the 
current error in $V_{cb}$ measurements. For our measurement of $V_{cb}$
the error contributed by uncertainty in $R_1$ and $R_2$ drops 
by a factor $\sim 4$ from that obtained
using the previous measurement from CLEO~\cite{CLEO}. The effect on the total 
uncertainty is to reduce it by $\sim 30\%$. 


A considerable improvement can also be obtained in
measurements of the lepton end point spectrum. The systematic error on the branching fraction 
for decays with a lepton in the momentum range $2.1-2.6 \gevc$ is reduced from $5\%$ to $1.7\%$ which 
corresponds to an improvement of $\sim 8\%$ in the total statistical plus systematic error.

In addition we have demonstrated useful approximations to the maximum
likelihood method that allow us to cope with the limited size of the Monte
Carlo samples available to modern high luminosity
experiments. We have also developed the procedures needed to evaluate the
corrections and additional uncertainties due to these approximations. These
methods are not unique to $D^*\ell\nu$ or $J/\psi K^*$, but could be applied 
to any analysis that needs
to cope with a complex multi-dimensional acceptance and resolution problem.

\section{Acknowledgments}
\label{sec:Acknowledgments}

The authors wish to thank Z.Ligeti and M.Neubert for very useful
discussions of the theoretical issues.


We are grateful for the 
extraordinary contributions of our \pep2\ colleagues in
achieving the excellent luminosity and machine conditions
that have made this work possible.
The success of this project also relies critically on the 
expertise and dedication of the computing organizations that 
support \babar.
The collaborating institutions wish to thank 
SLAC for its support and the kind hospitality extended to them. 
This work is supported by the
US Department of Energy
and National Science Foundation, the
Natural Sciences and Engineering Research Council (Canada),
Institute of High Energy Physics (China), the
Commissariat \`a l'Energie Atomique and
Institut National de Physique Nucl\'eaire et de Physique des Particules
(France), the
Bundesministerium f\"ur Bildung und Forschung and
Deutsche Forschungsgemeinschaft
(Germany), the
Istituto Nazionale di Fisica Nucleare (Italy),
the Foundation for Fundamental Research on Matter (The Netherlands),
the Research Council of Norway, the
Ministry of Science and Technology of the Russian Federation, and the
Particle Physics and Astronomy Research Council (United Kingdom). 
Individuals have received support from 
CONACyT (Mexico),
the A. P. Sloan Foundation, 
the Research Corporation,
and the Alexander von Humboldt Foundation.

\clearpage

{\noindent\bf Appendix A: Evaluation of additional errors}

Here we gather the equations used to evaluate the errors not included
in our likelihood fit. Specifically, we provide equations for the
errors due to the Monte Carlo integration of the REC method, that due
to the background subtraction of the DUBS method, and that due to the
fluctuations of the actual background initially left unaccounted for
when the DUBS method is used.

Full details and derivations can be found in the thesis of M.S. Gill\cite{Gill}.

\par\par\hfill\par
{\noindent\it REC Monte Carlo integration error}

The REC Monte Carlo integration error is given by:

\be
\label{eq:rec-err}
E_{rec}=E_{fit}DE^\dagger_{fit}
\ee
where $E_{fit}$ matrix is the error reported by the fitter.  $D$ is a
matrix of derivative sums.

$D$ is composed of three pieces. To lay them out we need to define some terms.

The sums of the weights $w_i\equiv F(\tilde x_i;\mu)/F(\tilde x_i;\mu_{mc})$ used in the integral:
\be
\label{eq:wsums}
W(\mu)=\sum{w_i}
\ee
where sums over MC events used to evaluate the normalization integral $N(\mu)$.

The vector of the derivatives  weights:
\be
\label{eq:dloglike}
\Delta_i\equiv\frac{dw_i}{d\mu}=
\left(\frac{\partial w_i}{\partial R_1},\frac{\partial w_i}{\partial R_2},\frac{\partial w_i}{\partial \rhosq}\right)
\ee
where $F_i\equiv F(\tilde x_i;\mu)$. The derivatives are computed numerically though an analytic 
compuation is possible.

The sum of weights squared:
\be
W_{sq}(\mu)\equiv\sum w^2_i
\ee

The $W$ derivative vector:
\be
 \frac{dW}{d\mu} =\sum \Delta_i
\ee
which also allows us to compute $\frac{dLnW}{d\mu}=\frac{1}{W}\frac{dW}{d\mu}$.

Weight $\times$ derivative vector:
\be
WdW=\sum  w_i\times \Delta_i
\ee
Derivative $\times$ Derivative matrix:
\be
dWdW=\sum \left(\Delta_i \Delta_i^{\dagger} \right) 
\ee

The three components are give by:
\be
D_1=\eta\times dWdW
\ee
\be
D_2=-\eta\times W_{sq}\left(\left(\frac{dW}{d\mu}\right)\times \left(\frac{dW}{d\mu}\right)^{\dagger}\right)
\ee
\be
D_3=-\eta\times \left(\left(\frac{dLnW}{d\mu}\right)WdW^{\dagger}+WdW\left(\frac{dLnW}{d\mu}\right)^{\dagger} \right) 
\ee
where $\eta=N_{signal}/W^2$ is a normalization factor. 

$D$ is just the sum of these three, i.e., $D=D_1+D_2+D_3$.

\par\hfill\par
{\noindent\it DUBS error}

The DUBS error can be computed from sums over the Monte Carlo sample
used in the background subtraction.  In this case the weights ($w_i$)
are those used to weight each type of background to obtain the correct
normalization as described in Sec. \ref{sec:back} (see eqs.
(\ref{eq:wpeak}) (\ref{eq:wcombo})).

In this we need the vector of the derivatives of the PDF $\frac{dLnF_i}{d\mu}$ from which we compute the `sensitivity' matrix $S$:

\be
\label{eq:sensitivity}
S_{dubs}=\sum w^2_i\left(\frac{dLnF_i}{d\mu} \right)\left(\frac{dLnF_i}{d\mu} \right)^{\dagger}
\ee

Given $S$ the DUBS error is just:
\be
E_{dubs}=E_{fit}S_{dubs}E^{\dagger}_{fit}
\ee

\par\hfill\par
{\noindent\it Background error}

We also use the DUBS sample to estimate
error from background in our signal sample. The result is analogous to
eq.~(\ref{eq:sensitivity}) accept that $w^2_i$ is replaced by
$w_i$. That is, we have:

\be
\label{eq:sensbk}
S_{back}=\sum w_i\left(\frac{dLnF_i}{d\mu} \right)\left(\frac{dLnF_i}{d\mu} \right)^{\dagger}
\ee
 and
\be
E_{back}=E_{fit}S_{back}E^{\dagger}_{fit}
\ee


\clearpage
\begin{thebibliography}{99}

\bibitem{NeubertPhysReport}
{M. Neubert, ``Heavy Quark Symmetry'', Physics Reports 259:245 (1994) [hep-ph/9306320].}

\bibitem{CLEO}
{ J.E. Duboscq et.al. (CLEO Collab.),  Phys. Rev. Lett. 76: 3898 (1996).}

\bibitem{isgurwise}
{N. Isgur and M.B. Wise, Phys. Lett. B237:527 (1990).}


\bibitem{Caprini}
{I. Caprini, L. Lellouch, M. Neubert, Nucl.Phys. B530:153-181 (1998) [hep-ph/9712417].}

\bibitem{LigetiGrinstein}
{B.Grinstein, Z.Ligeti,  Phys.Lett. B526:345-354 (2002) [hep-ph/0111392]. }

\bibitem{CloseWambach}
{F.Close, A. Wambach, Nucl.Phys. B412:169-180 (1994) [hep-ph/9307260]. }

\bibitem{ref:babar}
{The \babar\ Collaboration, B.\ Aubert {\em et al.},
Nucl.\ Instrum.\ Methods A479:1-116 (2002). }

\bibitem{babarVcb}
{BABAR, \vcb\ paper, Submitted Paper, ICHEP 2004 Conference Proceedings [hep-ex/0308027]}


\bibitem{babarsim}
{S. Agostinelli et al., Geant4 Collaboration, Nucl.\ Instrum.\ Methods A506:250-302 (2003).}

\bibitem{evtgen}
{D. Lange et al., Nucl.\ Instrum.\ Methods A462:152-155 (2001). }

\bibitem{pdg2002}
{Review of Particle Physics,
K. Hagiwara et al., Phys. Rev. D66, 010001 (2002), [http://pdg.lbl.gov] }

\bibitem{psikstar}{ Measurement of the $B \rar J /\psi K^*(892)$ Decay Amplitudes.
By the BABAR Collaboration, Phys.Rev.Lett.87:241801 (2001)
[hep-ex/0107049]. }

\bibitem{ichepLong}
{BaBar, ICHEP Form Factor Paper Submission Long Version  (2004) 
[http://www.slac.stanford.edu/\~\null msgill/FFdocs/ichepPaper.v11.ps]}

\bibitem{Gill}
{M. S. Gill, Ph.D. Thesis, University of California, Berkeley (UCB), (2004) 
[http://www.slac.stanford.edu/\~\null msgill/FFdocs/ffthesis.ps]}

\bibitem{Ryd}
{A. Ryd, Ph.D. Thesis, University of California, Santa Barbara (UCSB), (1997).}

\bibitem{Was}
{S. Jadach, B. Ward, Z. Was,
Comput.Phys.Commun.130:260-325 (2000).
}



\end{thebibliography}
\end{document}